# Formation of S0 galaxies through mergers

## Morphological properties:
## tidal relics, lenses, ovals, and other inner components

M. C. Eliche-Moral[1], C. Rodríguez-Pérez[2], A. Borlaff[1,3], M. Querejeta[4,5], and T. Tapia[6]

[1] Instituto de Astrofísica de Canarias, C/ Vía Láctea, E-38200, La Laguna, Tenerife, Spain, e-mail: carmen.eliche@gmail.com
[2] Departamento de Astrofísica y Ciencias de la Atmósfera, Universidad Complutense de Madrid, E-28040 Madrid, Spain
[3] Facultad de Física, Universidad de La Laguna, Avda. Astrofísico Fco. Sánchez s/n, 38200, La Laguna, Tenerife, Spain
[4] European Southern Observatory, Karl-Schwarzschild-Straße 2, D-85748 Garching, Germany
[5] Observatorio Astronómico Nacional (IGN), C/ Alfonso XII 3, E-28014 Madrid, Spain
[6] Instituto de Astronomía, Universidad Nacional Autónoma de México, BC-22800 Ensenada, Mexico

**ABSTRACT**

*Context.* Major mergers are popularly considered too destructive to produce the relaxed regular structures and the morphological inner components (ICs) usually observed in lenticular (S0) galaxies.
*Aims.* We aim to test if major mergers can produce remnants with realistic S0 morphologies.
*Methods.* We have selected a sample of relaxed discy remnants resulting from the dissipative merger simulations of the GalMer database and derived their properties mimicking the typical conditions of current observational data. We have compared their global morphologies, visual components, and merger relics in mock photometric images with their real counterparts.
*Results.* Only ~1–2 Gyr after the full merger, we find that: 1) many remnants (67 major and 29 minor events) present relaxed structures and typical S0 or E/S0 morphologies, for a wide variety of orbits and even in gas-poor cases. 2) Contrary to popular expectations, most of them do not exhibit any morphological traces of their past merger origin under typical observing conditions and at distances as nearby as 30 Mpc. 3) The merger relics are more persistent in minor mergers than in major ones for similar relaxing time periods. 4) No major-merger S0-like remnant develops a significant bar. 5) Nearly 58% of the major-merger S0 remnants host visually detectable ICs, such as embedded inner discs, rings, pseudo-rings, inner spirals, nuclear bars, and compact sources, very frequent in real S0s too. 6) All remnants contain a lens or oval, identically ubiquitous in local S0s. 7) These lenses and ovals do not come from bar dilution in major-merger cases, but are associated with stellar halos or embedded inner discs instead (thick or thin).
*Conclusions.* The relaxed morphologies, lenses, ovals, and other ICs of real S0s do not necessarily come from internal secular evolution, gas infall, or environmental mechanisms, as traditionally assumed, but they can result from major mergers as well.

## 1. Introduction

Lenticular or S0 galaxies are disc galaxies without spiral patterns and star formation (SF) spread in their discs, showing a diffuse and smooth appearance (see, e.g. Laurikainen et al. 2011; Erwin et al. 2012). Their number density has increased at the expense of spiral galaxies in clusters and groups since $z \sim 1$ (Dressler 1980; Dressler et al. 1997; Poggianti et al. 2009; D'Onofrio et al. 2015), becoming numerically dominant at $L \sim L^*$ at $z \sim 0$ over other morphological types in both environments (Marzke et al. 1994; Bernardi et al. 2010). Therefore, S0s must derive from spirals through mechanisms that have removed their gas and quenched the SF in their discs (e.g. Aragón-Salamanca et al. 2006; Laurikainen et al. 2011, L11 henceforth).

According to standard hierarchical scenarios, the mechanism governing the evolution of late-type galaxies (spirals) into early-type ones (specially, ellipticals and S0s) has essentially been galaxy mergers, mostly through major encounters (i.e. with mass ratios from 1:1 to 4:1) in the case of massive systems (Baugh et al. 1996; Somerville & Primack 1999; Maller et al.

2006; Hopkins et al. 2008; Kaviraj et al. 2009, 2014). Spirals would have then merged producing a S0 or an elliptical basically depending on the amount of gas re-accreted by the remnant from the environment. However, numerical simulations in the 1990s showed that major mergers are so catastrophic that they could hardly result in a system with an ordered disc, such as those of local S0s. Moreover, the remnants are expected to exhibit significant traces of their past merger origin for several giga-years after the merger, instead of the relaxed appearances that most present-day S0s have (L11; Laurikainen et al. 2006; Kaviraj 2014; Buta et al. 2015; Duc et al. 2015; Kormendy & Bender 2012, KB12 hereinafter).

S0s have very heterogeneous properties, many of which have been interpreted as evidence of the relative "gentleness" of the processes responsible of their formation, as opposed to those which would be expected if the bulk of this population had been formed through major mergers (Laurikainen et al. 2006; Emsellem et al. 2007; Laurikainen et al. 2009, 2010; Emsellem et al. 2011; Cappellari et al. 2011; Graham 2013; Vaghmare et al. 2013; Erwin et al. 2015; Mishra et al. 2017,

KB12). Consequently, many other processes have been proposed to explain how spirals can turn into S0s (see the review by Aguerri 2012). For instance, observational evidence of gas stripping in cluster galaxies has promoted the idea that environmental mechanisms (such as gas stripping, harassment, or strangulation) and passive disc fading must have been the main drivers of this transformation in clusters (Bekki et al. 2002; Crowl et al. 2005; Crowl & Kenney 2006, 2008; Vollmer et al. 2008, 2009, 2012; Rodríguez Del Pino et al. 2014). However, many authors point out that these processes may have been relevant only for intermediate-to-low mass galaxies in clusters, whereas mergers must have governed the evolution of massive ones (Ravikumar et al. 2006; Bernardi et al. 2011a,b; Barway et al. 2013; Vaghmare et al. 2015; Cerulo et al. 2017). Moreover, S0s do not preferentially live in clusters: at least half of them are found in groups in the field and some (very few) are isolated (Huchra & Geller 1982; Berlind et al. 2006; Crook et al. 2007; Wilman et al. 2009; Khim et al. 2015). This means that mechanisms different from cluster-related ones must also play a relevant role, at least to explain the S0 population in groups.

The morphological and structural properties of S0s point to internal secular evolution as another possibility. Nearly all S0s (∼97%) host lenses or ovals at their centres (Erwin & Sparke 2003; Laurikainen et al. 2013, L13 henceforth). This seems to be at the expense of strong bars, which are twice more common in their immediate spiral ancestors than in them (L13). Therefore, lenses and ovals are usually considered as relics of bar disruption and a sign of the relevance of internal secular evolution in the transformation of spirals into S0s (Buta et al. 2010). This view is also supported by the fact that many S0s also contain embedded inner discs with SF, nuclear bars, and central spirals patterns (Erwin & Sparke 2002; Emsellem et al. 2004; Falcón-Barroso et al. 2004; Kormendy & Kennicutt 2004; Gadotti et al. 2015; Sil'chenko 2015), often interpreted as byproducts of internal secular evolution, possibly linked to gas infall (Pfenniger & Norman 1990; Laurikainen et al. 2010, L10 hereinafter). Consequently, internal secular evolution and gas infall have become the main alternatives to environmental mechanisms, moving mergers to the backstage of current scenarios of S0 formation.

Nevertheless, many authors suggest that mergers and interactions may have triggered a more dramatic migration of spirals into S0s in groups than in clusters during the last 7 Gyr, mostly at high masses (Moran et al. 2007a,b; Silchenko & Afanasiev 2008; Wilman et al. 2008, 2009; Bekki & Couch 2011; Mazzei et al. 2014a,b; Hirv et al. 2017). Moreover, S0s in clusters are often located in groups and can exhibit clear merger relics (Rudick et al. 2009, 2010; Janowiecki et al. 2010; Mihos et al. 2013). These galaxies may have experienced a continuous processing through interactions and mergers as their group moves through the intracluster medium, simultaneously to environmental mechanisms (Berrier et al. 2006, 2009; De Lucia et al. 2012; Dressler et al. 2013; Weinzirl et al. 2014; Haines et al. 2015; Johnson et al. 2016). Signs of past mergers are found even in isolated S0s (Hargis et al. 2011).

Although extremely deep data reveal traces of past mergers around many local S0s, they are not ubiquitous and tend to be weak, so they are attributed more to satellite accretions, tidal interactions or environmental thickening than to major or intermediate encounters (Rudick et al. 2009; Janowiecki et al. 2010; Rudick et al. 2010; Thilker et al. 2010; Baillard et al. 2011; Duc et al. 2011, 2015, 2018; Kim et al. 2012; Mihos et al. 2013, KB12). Moreover, these tidal features are not exclusive of el-lipticals or S0s, but they are also frequent in galaxies of later types (Kaviraj 2010; Duc & Renaud 2013; Morales et al. 2018). Therefore, the role of mergers in the buildup of S0s is still a quite controversial topic.

As commented before, the most conventional argument used against the merger origin of some present-day S0s is that major mergers are too destructive to produce their ordered discs and relaxed morphologies. But many recent studies demonstrate the possibility of rebuilding discs in major mergers under certain conditions (see, e.g. Bekki 1998, 2001; Bournaud et al. 2005b, 2007b, 2011; Governato et al. 2007, 2009; Hopkins et al. 2009a,b; Bekki & Couch 2011; Hopkins et al. 2013; Athanassoula et al. 2016; Peschken 2017). However many authors are still reluctant to consider mergers as a feasible mechanism for S0 formation.

We have addressed this question by selecting a set of major and minor merger experiments from the GalMer database (Chilingarian et al. 2010) that result in relaxed remnants with significant disc components. We have visually analysed their morphological properties in realistic mock images (global appearance, sub-components, and tidal relics) to determine whether these remnants resemble real S0s or not, in order to discard or support the conventional catastrophic picture of major events. This paper is part of a series where many of the conventional arguments used against a possible major-merger origin of many S0s are demonstrated to be wrong (Borlaff et al. 2014; Querejeta et al. 2015a,b; Tapia et al. 2017, hereinafter B14, Q15a, Q15b, and T17, respectively). Here we specifically show that the following observational properties of S0s (usually attributed to internal secular evolution, disc fading or environmental processes) are compatible with a major-merger origin:

1. the relaxed and ordered discs of present-day S0s;
2. the lack of significant traces of past merging activity in most of them in relatively deep photometric images for relaxing times of just ∼1–2 Gyr;
3. the high percentage of ovals and lenses in real S0s, even without hosting or having hosted any large bars; and
4. the existence of embedded inner components (ICs), such as inner discs, rings, pseudo-rings, inner spirals, and nuclear bars at the centres of many S0s.

We describe the selection of the models in more detail than in previous papers in Sects. 2 and 3. Sections 4 and 5 explain the morphological classification performed to the selected remnants and how their global properties have been derived, mimicking current observational data. Section 6 describes the criteria and procedures adopted to identify the different morphological features. The analysis and discussion of the morphological properties of the remnants in comparison with their real analogues are exposed in Sects. 7 and 9. In particular, we analyse some trends of the remnant morphology with the initial conditions in Sect. 7.4. We discuss the limitations of the models in Sect. 8. The final conclusions are provided in Sect. 10. Several tables with information about the remnants have been included in Appendix A and the complete atlas of the 96 S0 and E/S0 remnants ("S0-like" remnants, hereinafter) is presented in Appendix B for the first time in this series.

A concordance cosmology has been assumed throughout the paper ($\Omega_M = 0.3$, $\Omega_\Lambda = 0.7$, $H_0 = 70$ km s$^{-1}$ Mpc$^{-1}$, Spergel et al. 2007). Magnitudes are provided in the Vega system.

Table 1: Baryonic mass ratios of `GalMer` merger experiments used here

| Major mergers | gE0 | gS0 | gSa | gSb | gSd |
|---|---|---|---|---|---|
| gE0 | 1:1 | – | 1.5:1 | 3:1 | 3:1 |
| gSa | – | – | 1:1 | 2:1 | 2:1 |
| gSb | – | – | – | 1:1 | 1:1 |
| gSd | – | – | – | – | 1:1 |
| Minor mergers | dE0 | dS0 | dSa | dSb | dSd |
| gS0 | 7:1 | 10:1 | 10:1 | 20:1 | 20:1 |

**Notes.** These data correspond to the `GalMer` experiments analysed in this series of articles, which were those available in the database on February, 2014: a total of 1002 experiments (876 major and 126 minor encounters).

## 2. GalMer merger simulations

`GalMer`[1] is a public database[2] containing more than 1100 binary N-body merger simulations at intermediate resolution (~250 000 particles per simulation), sampling a wide range of mass ratios, initial gas fractions, progenitor morphologies, and orbital conditions of the encounters, and accounting for the effects of gas and SF, beside gravitational ones (for a complete description, see Chilingarian et al. 2010). The experiments used progenitors of E0, S0, Sa, Sb, and Sd types in three different ranges of mass (giants, intermediate ones, and dwarves). The mass of the intermediate progenitors is lower than that of giants by approximately one half, while dwarves are less massive than giants by approximately one tenth. The mass ratios of the encounters thus range from 1:1 to 20:1, as indicated in Table 1. The experiments adopted different combinations of four initial values of the relative velocity between the progenitors (200, 300, 370, and 580 km s$^{-1}$) and three pericentre distances for the orbit (8, 16, and 24 kpc), resulting in twelve possible initial orbital configurations. There are also six possible inclinations between the discs and the orbital plane at $t = 0$ (0, 33, 45, 60, 75, and 90°) and two possibilities for the spin-orbit coupling (direct or retrograde). In all cases, the initial distance between the two galaxies was 100 kpc.

When we started this project, the database contained 876 major encounters involving giant galaxies of all types (except for gS0) and 126 minor events of dwarves of all types onto a gS0 progenitor. Therefore, our results are based on this set of 1002 simulations already available in the database on February, 2014. We will refer exclusively to them hereinafter.

The database contains the time evolution of each experiment for a total time period of ~3–3.5 Gyr, providing the physical status of each particle in the simulation every 50 Myr in FITS tables (position, velocity, mass, age, initial and current gas content and temperature in case of being hybrid, mean age and metallicity of the gaseous and stellar contents in them, etc.). The giant progenitors contain 120 000 particles in total (except for the gS0 progenitor, which has 480 000 particles) and the dwarves have 48 000, distributed among their components depending on their morphological type (see Table 2). These numbers account for collisionless stellar particles (we will also refer to them as old stellar particles or old stars), hybrid material (particles with an evolving fraction of its original mass in the form of gas and young stars) and dark matter ones. This means that we have a total of 240 000 particles in the major merger experiments and 528 000 in the minor merger ones.

The progenitor galaxies have a spherical non-rotating dark-matter halo and, depending on the morphological type, they can contain a non-rotating bulge, an old (collisionless) stellar disc, and a disc of hybrid particles, completely gaseous at the beginning of the simulation. Ellipticals do not contain any disc component; S0s have a bulge and an old stellar disc, but no initial gaseous disc; Sa and Sb types host a bulge, an old stellar disc, and a gaseous disc; and Sd's combine an old stellar disc and a hybrid one, but have no bulge. The bulge-to-disc ratios are 2.0 for the S0 progenitor, 0.7 in the Sa, and 0.4 for the Sb. The ratio of dark-to-baryonic matter depends on the progenitor, going from ~0.4 in the gE0 to ~3 in the gSd. The original haloes and bulges follow Plummer spherical profiles (Plummer 1911), while the stellar and gaseous discs are built according to the Miyamoto & Nagai (1975) density profile. The stellar disc has an Toomre parameter of $Q = 1.2$. Gas discs have two different values depending on the model ($Q = 0.3$ and $Q = 1.2$) and represent from 10% up to 30% of the total disc mass in the Sa–Sd progenitors, reproducing the typical gas fractions of these types in the present-day Universe. The sizes and masses per component of all progenitors are provided in Table 2 as a function of their type.

The gS0 progenitor is barred at the start of the simulation, whereas the other giant progenitors are not. Total stellar masses in the giant progenitors range ~0.5–1.5×10$^{11}$ M$_\odot$. Therefore, the final stellar masses of the remnants are ~1–3×10$^{11}$ M$_\odot$ in the major mergers and ~1.2–1.3×10$^{11}$ M$_\odot$ in the minor ones, depending on the efficiency of the star formation (SF) induced by the encounter and the masses of the progenitors. The mass of each particle then ranges ~3.5–20×10$^5$ M$_\odot$.

The simulations were computed with a TreeSPH technique, using the code by Semelin & Combes (2002). Gravitational forces are calculated using a hierarchical tree method (Barnes & Hut 1986) with a tolerance parameter $\theta = 0.7$, the resulting forces are then softened to a Plummer potential, and finally the code uses smooth particle hydrodynamics (SPH) to describe the gas and SF evolution (Lucy 1977; Gingold & Monaghan 1982). The softening length is $\epsilon = 280$ pc in the major merger experiments and $\epsilon = 200$ pc in the minor ones. The gas is considered as an isothermal component of temperature $T = 10^4$ K that turns into stars depending on the local gas concentration, according to the method by Mihos & Hernquist (1994b). The simulations also account for the effects of the stellar mass loss, the metallicity enrichment of the remaining gas with time, and the energy injection to the interstellar medium by SNe.

In the database, each merger experiment is referred as [$type_1$][$type_2$]o[$orbit$], where $type_1$ and $type_2$ represent the morphological type and size of the progenitors and $orbit$ corresponds to the identification number used in the `GalMer` database for each unique set of initial conditions (considering the complete orbital configuration, inclination, and spin-orbit coupling). In Table A.1, we have compiled the identification numbers and parameters corresponding to the orbital configurations in `GalMer` that result in fully merged and dynamically relaxed remnants in the experiments we have analysed (see Sect. 3). The progenitor labelled as 1 is always the most massive of the two merging galaxies in each simulation.

In many experiments, the progenitors merge into a single body just ~0.5–1 Gyr after the second pericentre passage, which is quickly virialised through violent relaxation (see Figs. 1 of B14 and Q15a for a sequence of snapshots showing the time evolution of a major and a minor merger, respectively). However, many remnants have not dynamically relaxed yet at the end of the simulation, some others are still at intermediate stages of the merger, while other models directly correspond to flybys. Therefore, we have firstly removed these cases from the sample to select only those remnants that could be comparable to real S0 galaxies (Sect. 3.1).

# 3. Selection of relaxed disc-like remnants

In order to restrict the sample only to those remnants that may really resemble local S0-like galaxies, we first selected the experiments from the `GalMer` database that generated apparently relaxed remnants with a relevant disc component in density maps (Sect. 3.1). Then, we checked that these remnants were structurally relaxed and nearly virialised (Sects. 3.2 and 3.3). The goal was to ensure a fair comparison of the detailed morphologies of these models and real data, so we could determine how consistent they really are (see Sect. 4).

## 3.1. Candidates to relaxed disc remnants

Three co-authors independently made a first selection of the models that, at the end of the simulation, visually resulted in remnants that: 1) had all their stellar material bounded into one single body, 2) exhibited apparent relaxed global morphologies, and 3) hosted a significant disc component in projected density maps. With the first two criteria, the flybys and galaxies still merging at the end of the simulation were excluded. The last criterion rejected many remnants which would obviously exhibit elliptical morphology in mock photometric images.

We used the viewer available in the `GalMer` database for this purpose, which allows the visualisation of projected density maps of the different components of the merging galaxies for different snapshots, point of views, and distances to the object. In this pre-selection, we included any models chosen by at least one co-author according to the criteria exposed before. From the initial 1002 models available in the database, we finally pre-selected 215 major and 72 minor merger simulations as candidates to result in S0 or E/S0 remnants (287 in total). They are listed in Table A.2.

We emphasise that these 287 remnants had a stellar disc component that could be identified in the projected density maps. Nevertheless, these maps are not equivalent to realistic photometric images. In fact, many of these discs are lost in mock photometric images (see Sect. 4). This means that all remnants in Table A.2 have a discy component, even though some of them are finally classified as ellipticals attending to their morphologies in the images.

## 3.2. Fully merged and structurally relaxed remnants

Before classifying them morphologically, we determined which remnants from our first selection could be considered as fully merged and structurally relaxed according to a set of qualitative and quantitative criteria, which differed depending on whether the remnant was from a major or a minor merger.

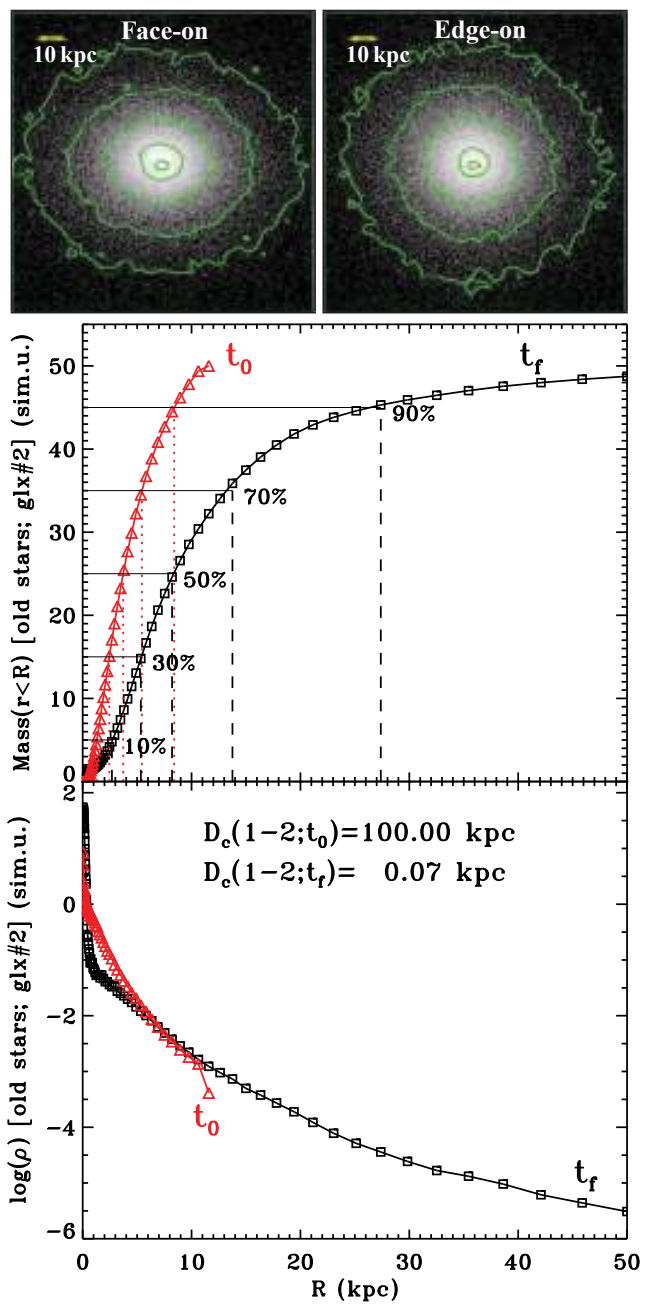

Fig. 1: Diagnostics used to confirm that the remnant of model gSagSao23 is fully merged and structurally relaxed (Sect. 3.2). We represent the radial profiles of the accumulated mass contained within each radius (*middle panel*) and volume density (*bottom panel*) of the old stellar material of progenitor 2 in the final remnant of the model. We compare the profiles for the initial and final times of the simulations (*red triangles* and *black squares*, respectively), centred in its own mass distribution at $t_0 = 0$ and in the mass centroid of progenitor 1 in the final time, $t_f$. The radii enclosing different percentages of the old stellar mass of progenitor 2 are marked in the top panel for each time (*dotted* and *dashed vertical lines*, respectively). The final profiles of both physical quantities indicate that the stellar content of progenitor 2 is well mixed with that of progenitor 1 in the final remnant. Moreover, the final distance between the centroids of the stellar material of each progenitor, $D_c$, is lower than the softening length of the simulations at the final time (see the bottom panel). This remnant is an elliptical, according to its morphology in its mock *V*-band images (top panels).

Table 2: Properties and number of particles used to model the progenitor galaxies in GalMer experiments

| Component | Parameter | gE0 | gS0 | gSa | gSb | gSd | dE0 | dS0 | dSa | dSb | dSd |
|---|---|---|---|---|---|---|---|---|---|---|---|
| Bulge | $M_B$ [$2.3 \times 10^9 M_\odot$] | 70 | 10 | 10 | 5 | 0 | 7 | 1 | 1 | 0.5 | 0 |
| | $r_B$ [kpc] | 4 | 2 | 2 | 1 | ... | 1.3 | 0.6 | 0.6 | 0.3 | ... |
| Halo | $M_H$ [$2.3 \times 10^9 M_\odot$] | 30 | 50 | 50 | 75 | 75 | 3 | 5 | 5 | 7.5 | 7.5 |
| | $r_H$ [kpc] | 7 | 10 | 10 | 12 | 15 | 2.2 | 3.2 | 3.2 | 3.8 | 4.7 |
| Disc | $M_\star$ [$2.3 \times 10^9 M_\odot$] | 0 | 40 | 40 | 20 | 25 | 0 | 4 | 4 | 2 | 2.5 |
| | $M_{g.D}/M_\star$ | ... | ... | 0.1 | 0.2 | 0.3 | 0 | 0 | 0.1 | 0.2 | 0.3 |
| | $a_{\star,D}$ [kpc] | ... | 4 | 4 | 5 | 6 | ... | 1.3 | 1.3 | 1.6 | 1.9 |
| | $h_{\star,D}$ [kpc] | ... | 0.5 | 0.5 | 0.5 | 0.5 | ... | 0.16 | 0.16 | 0.16 | 0.16 |
| | $a_{g.D}$ [kpc] | ... | ... | 5 | 6 | 7 | ... | ... | 1.6 | 1.9 | 2.2 |
| | $h_{g.D}$ [kpc] | ... | ... | 0.2 | 0.2 | 0.2 | ... | ... | 0.06 | 0.06 | 0.06 |
| No. particles | $N_g$ | ... | ... | 20 000 | 40 000 | 60 000 | ... | ... | 8 000 | 16 000 | 24 000 |
| | $N_\star$ | 80 000 | 80 000 | 60 000 | 40 000 | 20 000 | 32 000 | 32 000 | 24 000 | 16 000 | 8 000 |
| | $N_{DM}$ | 40 000 | 40 000 | 40 000 | 40 000 | 40 000 | 16 000 | 16 000 | 16 000 | 16 000 | 16 000 |

**Notes.** The masses and effective radii of the bulges in the progenitor galaxies (in all types, except for Sd's) are referred as $M_B$ and $r_B$. $M_H$ and $r_H$ represent analogous properties for their dark matter halos. $M_{\star,D}$, $a_{\star,D}$, and $h_{\star,D}$ are the masses, effective radii, and scale-heights of the discs made of collisionless stellar particles (in progenitors from S0 to Sd types). From Sa to Sd types, the progenitors also contain discs of hybrid particles, totally gaseous at the start of the simulations, with analogous properties to stellar ones, but denoted as $M_{g.D}$, $a_{g.D}$, and $h_{g.D}$. The bottom row of the Table indicates the number of particles in each component: $N_{DM}$ corresponds to the dark matter halo, $N_\star$ refers to the collisionless stellar particles distributed in the bulge and disc, and $N_g$ to the hybrid particles in the initial gaseous disc.

### 3.2.1. Major mergers

Concerning major mergers, we forced the remnants to fulfil the following conditions to consider them as totally merged and structurally relaxed:

1. The distance between the centroids of the mass distributions of the stellar material originally belonging to each progenitor must be lower than the softening length ($\epsilon \sim 0.28$ kpc) in the final remnant.
2. The mass distribution of progenitor 2 must be centred in that of progenitor 1 at the end of the simulation and must exhibit a smooth and regular spatial mass distribution.
3. The original mass profile of progenitor 2 must have been significantly redistributed in the remnant, according to the dynamical response expected for a system undergoing violent relaxation, this is the profile must have experienced an expansion of the external material towards outer radii and a compression in the inner regions.

The first condition ensured that the material of the two progenitors has coalesced and, in particular, that the original bulges, which are denser and thus take more time to be disrupted than the discs, had been completely destroyed by the tidal forces. The remnants that did not fulfil it were classified as on-going mergers. The second and third conditions guaranteed that the material of both galaxies had been completely mixed in the remnant and the global structure was sufficiently relaxed. These conditions were checked by computing the radial profiles of the volumetric mass density and the accumulated mass at each radius of the stellar material coming from progenitor 2 in the final remnant, using the mass centroid of the stellar material originally in progenitor 1 as the origin of coordinates. If the profiles visually exhibited bumps at a given radius, did not peak in the centre, or had not been widely redistributed in the remnant, one of these two conditions or both were not fulfilled, indicating that the two galaxies were still merging or undergoing violent relaxation. On the contrary, a major-merger remnant was fully merged and structurally relaxed if it obeyed the three previous criteria (an example is provided in Fig. 1).

The relaxation level of the pre-selected remnants resulting from major mergers is indicated in Table A.2. We finally identified 175 major-merger remnants (out of the original 215) as fully merged and structurally relaxed, according to the criteria described above.

### 3.2.2. Minor mergers

In minor mergers, the dwarf material is acquired by the gS0 progenitor during the successive approaches of the satellite in the orbit, so it is deposited following the orbital geometry, many times giving rise to slightly non-axisymmetric structures in the remnant. This made the remnant of several minor mergers not to fulfil the first condition commented before, despite the satellite had been fully accreted by the main progenitor. Moreover, the satellite cores were very resistant to be fully destroyed in the minor merger simulations. So, we detected several remnants of minor mergers that, obeying the criteria imposed to major ones, still had the satellite bulge undisrupted and orbiting around the remnant centre. We therefore had to adapt the previous conditions to minor merger cases as follows:

1. The distance between the mass centroids of the stellar particle distributions initially belonging to each progenitor in the final remnant should be lower than the original characteristic scale of the satellite.
2. Analogously to major mergers, the volumetric mass distribution of the satellite at the end of the simulation had to be centred in the mass centroid of the main gS0 progenitor and have a smooth and regular radial profile.
3. The final remnant should contain less than 10% of the total mass of the accreted satellite within a radius equal to the original characteristic scale of the satellite.

For dE satellites, we assumed the bulge effective radius as their characteristic scale ($r_B$ in Table 2). For dS0 or dSa–dSd ones, this scale was considered as the disc effective radius instead ($a_{\star,D}$, see Table 2). The first condition should ensure that the satellite was not orbiting the remnant centre, but it had been mostly disrupted. Although this was true for most cases, we realised that some models that satisfied this condition were still on-going the merger. In them, the simulation had been casually stopped at a moment in which the satellite core was flying by the centre of the gS0, but none of them were fully merged systems in which this core had simply been deposited undisrupted at the remnant centre. However, the fulfilment of the other two conditions rejected these cases from the sample. They ensured that the satellite core had really been destroyed and that its material had been redistributed in the main gS0 progenitor, because the percentage of accumulated mass in the characteristic radius of the original satellites was much higher than just 10%. The remnants of minor events that fulfilled none of these three criteria were on-going mergers, those that obeyed them partially were merged, but structurally unrelaxed systems, and those that fulfilled the three conditions were considered fully merged and structurally relaxed. Figure 2 illustrates the three cases.

The relaxation level of the pre-selected remnants resulting from gS0+dwarf encounters is also indicated in Table A.2. We finally identified 29 minor-merger ones (out of the initial 72) as fully merged and structurally relaxed, according to the previous criteria.

### 3.3. Dynamically relaxed remnants

The criteria exposed in Sect. 3.2 selected remnants with an apparent relaxed structure, but not necessarily virialised. We checked which remnants out of the 204 selected models were near dynamical equilibrium (175 major and 29 minor mergers).

In practice, a remnant is considered near dynamical equilibrium if it has relaxed for a time period ($T_{\text{relax}}$) much longer than the orbital period of a particle ($T_{\text{orb}}$) at its most external radius ($R_{\text{max}}$). The relaxing time is defined as the period elapsed between the time of the full merger (i.e. the moment when the mass centroids of the two progenitors nearly overlap in the simulation) and the final computed time (an example is shown in Appendix B). For major mergers, we conservatively considered that the remnants in which $T_{\text{relax}}$ exceeded $T_{\text{orb}}$ at $R_{\text{max}}$ at least by a factor of ∼4 were in dynamical equilibrium.

For minor mergers, we had to reduce this factor to ∼2 because they take longer to merge than major ones. A criterion as restrictive as in major mergers would have discarded most of the minor-merger remnants as non virialised. Taking into account that they imprint much less damage to the main progenitor structure than major ones, we considered that a lower relaxation level in minor mergers is acceptable for our analysis. The lower relaxing times allowed in minor mergers than in major ones obviously contribute to the higher fraction of S0 remnants with merger relics found in the minor-merger sample than in the major-merger one (Sect. 7.3). However, independently of this fact, we have also found that tidal features tend to be more persistent in minor events than in major ones even for similar relaxing time periods (of ∼1 Gyr, see the same Section).

We estimated the maximum visual radii of the remnants using their artificial face-on photometric images in the $K$ band (see Sect. 4). Approximating the total mass enclosed within $r \leq R_{\text{max}}$ by the total mass of the object ($M$), the orbital period of a particle at its outermost visual radius is approximately: $T_{\text{orb}} \sim 2\pi [R_{\text{max}}/(GM)]^{1/2}$, where $G$ represents the gravitational constant and $M$ is the sum of the total masses of the two progenitor galaxies in each case, including baryonic and dark matter contents. In the GalMer experiments, $M$ only depends on the morphological types of the progenitors.

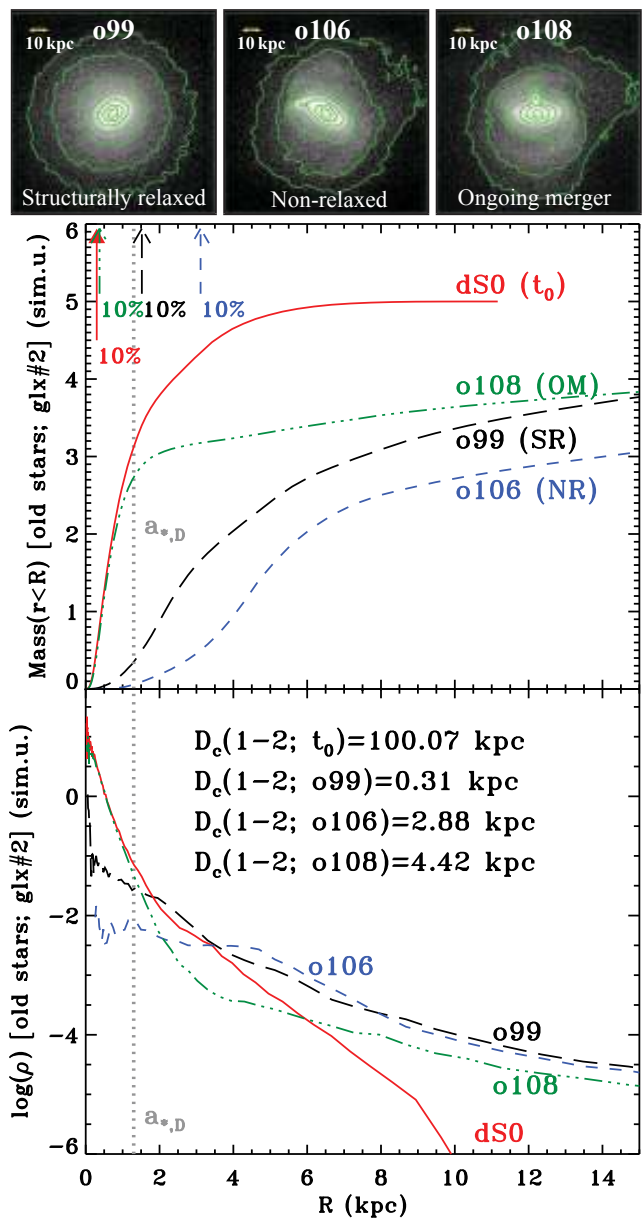

**Fig. 2:** Radial profiles of the accumulated mass within each radius (*middle panel*) and volume density (*bottom panel*) of the old stellar content of the dwarf in the remnants of three gS0+dS0 minor-merger experiments with different orbits, compared to the original profile of the dS0 (*red solid line*). The remnants present different relaxation levels according to the criteria exposed in Sect. 3.2: an ongoing merger (orbit #108, *green dotted-dashed line*), a non-relaxed system (#106, *blue short dashed line*), and a fully merged and structurally relaxed remnant (#99, *black long dashed line*). The *arrows* in the top panel indicate the radius containing 10% of the initial stellar mass of the satellite in each case (see the legend). The *vertical grey dotted line* indicates the disc scalelength of the dS0 ($a_{\star,D}$). Although the accumulated mass at this radius is <10% in orbits #106 and #99, the distance between the centroids of the stellar content of #106 is > $a_{\star,D}$ in the remnants of #106, so it is not relaxed (bottom panel). However, it is lower in the remnant of the orbit #99, thus being structurally relaxed. In #108, neither the mass accumulated in $a_{\star,D}$ is <10% nor the distance between the centroids is < $a_{\star,D}$, hence it is an on-going merger. This classification is consistent with their morphologies in their face-on *V*-band mock images (top panels).

For simplicity, we considered the highest value of all $R_{max}$ values estimated for the remnants resulting from the same couple of progenitor types as the characteristic value of the whole family of models. This value provides an upper limit for the $T_{orb}$ values of all remnants coming from the same couple of progenitor types, because the final mass is the same in them all, and therefore it also defines a conservative minimum value for the relaxing time required to be considered in dynamical equilibrium for the whole family. The characteristic values of $R_{max}$ and $T_{orb}$ obtained for each pair of progenitor types are listed in Table 3. The last column of the Table indicates the minimum relaxing times to be considered dynamically relaxed for each combination of progenitor types: $T_{relax,min} \sim c \times T_{orb}$, being $c = 4$ in major mergers and $c = 2$ in minor ones.

Additionally, we have estimated the relaxing times undergone by each experiment in our sample of 204 merged and structurally relaxed remnants (see Table A.3). They ranged $\sim$0.3–2 Gyr, so comparing them with the corresponding $T_{relax,min}$ values in Table 3, we finally determined that 173 out of the 175 major-merger remnants and all the selected minor mergers were near dynamical equilibrium. The models that had not relaxed enough time according to their $T_{relax,min}$ values were removed from the sample (models gE0gSdo71 and gSagSdo74). We finally obtained a sample of 202 dynamically relaxed remnants (173 major and 29 minor merger models). They distribute nearly equally among prograde and retrograde encounters ($\sim$49% vs. $\sim$51%, respectively). We will refer to these 202 dynamically relaxed remnants simply as relaxed remnants henceforth, unless otherwise noted.

## 4. Remnants with realistic S0 and E/S0 morphologies

The majority of N-body studies analyse the morphology of merger remnants using projected density maps, plots of projected particle positions, or assume a constant mass-to-light ratio ($M/L$) to transform density maps into surface brightness ones. The two formers saturate in dense regions, masking substructures, and none accounts for neither the difference between mass and luminosity in particles formed at different epochs, nor the observational biases intrinsic to real data, which can significantly affect the appearance of low-density regions. In fact, many disc components detected in the projected density maps of the remnants in Sect. 3.1 become undetectable under realistic observing conditions, meaning that an observer would identify them as ellipticals, instead of S0s. A fair visual comparison of the remnants with real S0s thus requires the removal of these elliptical cases from our sample.

We have performed the morphological classification of the remnants visually using synthetic photometric images, precisely to mimic the effects of both the complex stellar populations of the remnants and the observational biases. Here we describe how we have created these mock photometric images (assuming the typical observing conditions of current ground-based local surveys, see Sect. 4.1) and performed the visual morphological classification attending to them (Sect. 4.2).

### 4.1. Mock photometric images

#### 4.1.1. Description of the procedure

The photometric images of the remnants were created in five photometric broad bands ($B$, $V$, $R$, $I$, and $K$), assuming that they

Table 3: Characteristic minimum time periods for dynamical relaxation of the remnants

| Models (1) | $R_{max}$ [kpc] (2) | $M_{dyn}$ [$2.25 \times 10^9 M_\odot$] (3) | $T_{orb}$ [Gyr] (4) | $T_{relax,min}$ [Gyr] (5) |
|---|---|---|---|---|
| **Major mergers** | | | | |
| gE0+gSa | 15.8 | 204.0 | 0.11 | 0.44 |
| gE0+gSb | 21.6 | 204.0 | 0.12 | 0.48 |
| gE0+gSd | 14.8 | 207.5 | 0.06 | 0.24 |
| gSa+gSa | 22.5 | 208.0 | 0.16 | 0.64 |
| gSa+gSb | 26.4 | 208.0 | 0.14 | 0.56 |
| gSa+gSd | 27.2 | 211.5 | 0.16 | 0.64 |
| gSb+gSb | 22.8 | 208.0 | 0.07 | 0.28 |
| gSb+gSd | 26.5 | 211.5 | 0.10 | 0.40 |
| gSd+gSd | 22.4 | 215.0 | 0.12 | 0.48 |
| **Minor mergers** | | | | |
| gS0+dE0 | 18.8 | 110.0 | 0.24 | 0.48 |
| gS0+dS0 | 18.8 | 110.0 | 0.24 | 0.48 |
| gS0+dSa | 16.8 | 110.4 | 0.21 | 0.42 |
| gS0+dSb | 17.2 | 110.4 | 0.21 | 0.42 |
| gS0+dSd | 15.6 | 110.75 | 0.18 | 0.36 |

**Notes.** *Columns:*: (1) family of merger models using this combination of morphological types of the progenitors. (2) Maximum value of the external radii of the remnants resulting from this family of models, visually determined from the mock $K$-band images of the stellar remnants. (3) Total dynamical mass of the remnants resulting from the merger of these progenitor types (dark matter + old stars + hybrid material), in simulation units (equal to $2.25 \times 10^9 M_\odot$). (4) Characteristic orbital time period of a particle at $R = R_{max}$ in these remnants. (5) Minimum relaxing time period required for the remnants of this family of mergers to be considered in dynamical equilibrium (see Sect. 3.3).

were located at a given distance. S0s may be confused with ellipticals at low inclinations (Wilman et al. 2009), so we have simulated images in face-on and edge-on views of each remnant to facilitate the identification of disc components. One example of the set of images for each remnant is shown in Appendix B.

The face-on view of a remnant was defined as the line of sight in the direction of the total angular momentum of its baryonic content, while the edge-on view was chosen as the intersection between the plane perpendicular to it and the XY plane of the original coordinate system. In some experiments with very inclined orbits, a secondary disc formed, inclined with respect to the main remnant disc. In these cases, the total angular momentum of the remnant is deviated towards a direction which is not strictly perpendicular to the plane of the main remnant disc. These cases are easily recognizable because the main remnant disc does not exhibit an inclination of 90º in the edge-on view defined here (see some examples in Sect. 7.2.2).

To transform projected density maps into realistic synthetic photometric images, it is required a proper mass-to-light conversion and the insertion of observational biases inherent to real data, such as cosmological dimming, limiting magnitude, spatial resolution, and atmospheric blurring. Considering the information of each stellar particle (collisionless or hybrid) in the final snapshot of the remnant, their locations are projected onto a spatial grid on the sky orientated according to a selected line of sight to the remnant and taking into account the scaling corresponding to the distance assumed for it (Sect. 4.1.4). Intrinsic physi-cal lengths in the projected density maps are converted into projected angular scales in the sky, assuming a concordance cosmology. The projected angular grid already accounts for the assumed spatial resolution element (Sect. 4.1.3). The stellar mass content of each particle in a given pixel of the grid is converted into luminosity in the selected band using the $M/L$ value on the basis of a set of stellar population models, as described in Sect. 4.1.2. The total luminosity coming from all stellar particles that project in a given pixel is turned into flux. The cosmological dimming corresponding to the assumed distance is then introduced, photon and sky noises accordingly to the adopted limiting magnitude are inserted, and the photometric image is finally convolved by the assumed PSF (Sect. 4.1.3). For simplicity, we assumed a gain of one electron per ADU.

Dust extinction effects were not included in the images to avoid adopting any assumptions about its geometrical distribution and physical properties, as well as to elude the numerical problems usually entailed by recipes dealing with it (Jonsson et al. 2010; Bekki 2013a,b). Moreover, neglecting dust effects is justified in this case, because S0s have been traditionally considered devoid of dust and gas (e.g. Falcón-Barroso et al. 2006; Williams et al. 2010). In any case, the hybrid particles in the remnants, which would contain most of the dust resulting from the merger-induced SF, tend to accumulate at their innermost regions, so we expect that dust extinction does not significantly affect the global morphology of the remnants, except at the cores.

### 4.1.2. Mass-to-light conversion

For simplicity, we assumed that all collisionless stellar particles in each remnant had the same age, independently of the progenitor they originally belonged to. Assuming that the remnants are observed at $z \sim 0$ (Sect. 4.1.4), this age was set to ~10 Gyr, since this is the average age of old stars in the discs of present-day S0s (Finkelman et al. 2010; Sil'chenko et al. 2012; Sil'chenko 2013). This age nearly coincides with the lookback time corresponding to the peak of the SF history (SFH) of the Universe ($z \sim 2$, see Hopkins & Beacom 2006; Cooper et al. 2008; Reddy et al. 2008; Shi et al. 2009; Madau & Dickinson 2014).

We assigned a SFH to each collisionless particle depending on the morphological type of its original progenitor, adopting parametrisations for each type based on observations: a simple burst in E and S0 progenitors, exponentially-decaying SFHs in Sa and Sb ones with e-folding timescales $\tau = 4$ and 7 Gyr respectively, and a constant SFR in Sd cases. The complete parametrisations are provided in Table 2 in Eliche-Moral et al. (2010). The mass of these old stellar particles was considered to have evolved according to these SFHs since their formation epoch, in other words since a lookback time of ~10 Gyr in all cases, until the epoch at which the merger was considered to have started (i.e. at a lookback time of ~3–3.5 Gyr, depending on the model, see Table A.3).

Since the merger starts, the SF in the galaxies was considered to decouple from the old stellar particles and become exclusively linked to the hybrid ones, which are the ones tracing the gas evolution during the simulation. Therefore, the standard SFH that each old stellar particle has been experiencing since its formation at high redshift is truncated at the cosmological time assumed as the starting point of the merger simulation, and thus it is considered to have evolved passively during the merger until it finishes (at $z = 0$). With this procedure, we were accounting for both the SFH experienced by old stellar particles in the progenitor galaxies *before the merger* and their passive evolution *during*

Table 4: Observing conditions adopted for the artificial photometric images of the remnants

| Band | $\mu_{lim}$ [mag arcsec$^{-2}$] | $(S/N)_{lim}$ | FWHM [arcsec] | References |
|------|------|------|------|------|
| (1) | (2) | (3) | (4) | (5) |
| $B$ | 26.0 | 5 | 0.7 | 1 |
| $V$ | 26.0 | 5 | 0.5 | 2 |
| $R$ | 27.5 | 5 | 0.7 | 3, 4 |
| $I$ | 26.0 | 5 | 0.7 | 1 |
| $K$ | 22.0 | 3 | 0.7 | 5 |

**Notes.** *Col. 1*: photometric band. *Col. 2*: assumed limiting surface brightness in mag arcsec$^{-2}$ (Vega system). *Col. 3*: signal-to-noise ($S/N$) ratio corresponding to the limiting surface brightness. *Col. 4*: full width at half maximum of the assumed PSF in arcsec. *Col. 5*: reference observational studies.

**References.** (1) SDSS, Abazajian et al. (2009); (2) KB12; (3) Erwin et al. (2008); (4) Gutiérrez et al. (2011); (5) NIRS0S, L11.

*it*. The evolution of $M/L$ in each band for these truncated SFHs was estimated using the stellar population synthesis models by Bruzual & Charlot (2003), adopting a Chabrier initial mass function (Chabrier 2003) and the Padova 1994 evolutionary tracks (Bertelli et al. 1994).

Hybrid particles do not play any role in the SFH of the progenitors before the merger starts, because they are totally gaseous at the beginning of the encounter. During the merger, part of their initial masses is transformed into stars at each step of the simulation according to certain recipes (Sect. 2). The SFH undergone by each hybrid particle can be very complex, because it depends on local conditions at each time, but it is basically concentrated in two short SF peaks which occur soon after the first pericentre passage and the full merger (which nearly coincides with the second pericentre passage most times), with most SF taking place in the last one in reality (see Di Matteo et al. 2007; di Matteo et al. 2008; Lotz et al. 2008, and Eliche-Moral et al., in prep.). Therefore, we approximated the SFH undergone by each hybrid particle (basically concentrated in a peak after the full merger) by a simple stellar population (SSP) model, assuming for it an age and metallicity equal to the average age and metallicity of the stellar content of the particle at the end of the simulation. We also used the models by Bruzual & Charlot (2003) for computing the $M/L$ evolution of the SSP models for each hybrid particle, assuming a Chabrier IMF and the Padova prescription too.

### 4.1.3. Observing conditions

We adopted observing conditions in each band similar to those exhibited by the data of current ground-based observational surveys, in particular, concerning the limiting magnitudes, spatial resolutions, and atmospheric blurring. This guaranteed that the artificial images mimicked the conditions of realistic data, allowing a direct comparison between the real S0s and our S0-like remnants, as done in other papers of this series (B14; Q15a,b; T17). Table 4 indicates the limiting surface brightness at certain signal-to-noise ratios ($S/N$) and FWHM values of the seeing assumed for each band, as well as the observational studies taken as reference for selecting them. The size of the pixels was set to half of the FWHM of the PSF in each case.

For the seeing conditions adopted in these photometric images, the appearance of the outer disc in the remnants looks grainy because of the presence of extremely bright individual particles in certain pixels, whereas adjacent pixels that are empty of stellar particles remain dark (see some examples in Appendix B). This is because each particle represents a stellar mass of $\sim 3.5$–$20 \times 10^5 M_\odot$, equivalent to the typical mass of a globular cluster (Sect. 2). This discontinuous structure of the outer disc makes difficult the identification of these particles as a unique, bounded structural component by automatic 2D fitting algorithms. Nevertheless, a 1D azimuthal average of the light distribution provides realistic surface brightness profiles for the disc, meaning that, although the models have obvious limitations due to their discrete nature, the global structure of the remnants is realistic (for more details, see Fig. 3 in Q15a).

### 4.1.4. Distance to the remnants

The distance to the remnants was set to $D = 30$ Mpc in all cases because of several reasons. Firstly, this was the average distance of the S0 galaxies in the NIRS0S survey[3] (L11), which we chose as the observational reference sample in the $K$ band (Q15a). Secondly, this distance was similar to the furthest one exhibited by the anti-truncated S0s of the samples compiled by Erwin et al. (2008) and Gutiérrez et al. (2011) in the $R$ band, which we used for comparison with the anti-truncated discs found within our S0-like remnants (B14). By assuming the furthest distance in this case, we were adopting the worst observing conditions of the reference data for our simulations. This was relevant because anti-truncations are features located at the disc outskirts, in the lowest $S/N$ regions of the data, so we wanted to guarantee that any similarities found between real and simulated anti-truncations existed despite the limitations intrinsic to observations. And finally, the FWHM used in $R$ and $K$ bands was equivalent to an intrinsic spatial resolution of $\sim 100$ pc for this distance (Table 4), enough for performing the morphological classification of galaxies with optical sizes $R_{25}(V) \sim 10$–$25$ kpc, like ours. However, we point out that the distances assumed in other papers of this series may differ, depending on the observational sample used as reference (see, e.g. Q15b).

### 4.2. Morphological classification

Although several automatic methods for morphological classification of galaxies have been developed in the last years (Schawinski et al. 2007; Jogee et al. 2009; Darg et al. 2010; Shamir 2011), visual classification is still used in relatively small samples because it is very efficient for identifying faint structures (Abraham et al. 1999; Bamford et al. 2009, L11). Therefore, we visually identified the remnants in our sample with S0 or E/S0 morphology.

Five co-authors independently classified the morphology of the 202 relaxed remnants attending to their face-on and edge-on mock photometric images in the five broad bands commented previously. We assigned traditional morphological types: elliptical (E) if no disc was detectable in the images; E/S0 if there was a disc besides the bulge component, but it was sub-dominant compared to the spheroidal component; lenticular (S0) if the disc dominated over the bulge and it had no spiral patterns or SF knots; and spiral (S) if the disc clearly had spiral arms. The visual distinction between S0 and E/S0 is somewhat arbitrary, because it depends on whether the classifiers subjectively considered that the spheroid dominated the galaxy morphology over the disc or not. Additionally, we identified relevant bars in the remnants. The final morphological type of each object was defined as the median type of the five independent classifications performed for it. In $\sim 85\%$ of the cases, the five classifiers agreed, meaning that the classification is quite homogeneous and robust.

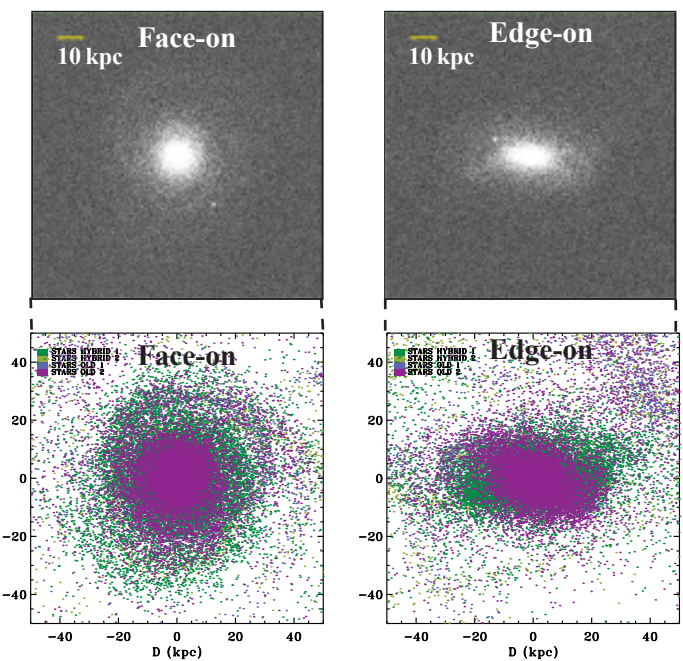

Fig. 3: Comparison of the morphology of the final remnant of model gSbgSdo71 (E/S0) in its $V$-band mock images (*top*) and its maps of projected particle positions (*bottom*), for the face-on and edge-on views. We have assumed a distance of $D = 30$ Mpc, $\mu_{lim}(V) = 26$ mag arcsec$^{-2}$ (Vega) for $S/N = 5$, and a FWHM= 0.5″ for the photometric images (Table 4). The selected grayscale emphasises the sky noise of the images. In the maps, different types of stellar particles in the remnant are plotted using different colours (*dark green*: hybrid particles, progenitor 1; *blue*: old stars, progenitor 1; *light green*: hybrid particles, progenitor 2; *purple*: old stars, progenitor 2). The field of view (FoV) is $100 \times 100$ kpc$^2$ in all frames. The horizontal line at the top left corner of the top panels represents 10 kpc. Notice how faint tidal structures at the outskirts of the maps are strongly diluted in the photometric images when observational effects are realistically accounted for.

The surface brightness profiles in the five bands were used in limiting cases for discriminating whether the remnant had a significant disc component or not. In the remnants with huge spheroidal components, sometimes it was not clear from the edge-on images whether the outer region was dominated by the outwards extrapolation of the central bulge or by a faint disc. By looking to the surface brightness profiles, we could distinguish if the outer section of the profile decreased radially following an exponential law (thus being a disc) or it obeyed the $R^{1/n}$ law defined by the central spheroid instead (see Eliche-Moral, in prep.).

In order to illustrate the relevance of accounting for the observational limitations when comparing the visual morphologies of these remnants with real S0s, we have represented the *V*-band images and the corresponding maps of projected positions of the stellar particles of one remnant finally classified as E/S0 in Fig. 3. Its morphology drastically changes when observational effects are considered, even for distances as nearby and magnitudes as deep as those assumed here. The outer disc of the remnant is basically lost in the mock images, as well as the faint external features, such as the tidal tails and ripples, which are clearly visible in the particle-position maps instead. In fact, many of the selected relaxed remnants with disc components visible in their density maps are finally classified as ellipticals, because their discs become visually undetectable in these realistic mock images. Moreover, the global morphology of the remnant appears much more relaxed in the photometric images in the Figure. Therefore, our procedure guarantees that we have selected only the remnants that would be visually identified as S0-like galaxies by any observers under typical observing conditions of current ground-based photometric data.

The 29 minor mergers produced a remnant of S0 type attending to their photometric images, expected since the main progenitor was already a gS0 in all cases and the mass ratios were higher than 7:1 (Table 1). Concerning the 173 major-merger remnants, we found 106 ellipticals (61.3%), 25 E/S0's (14.4%), and 42 S0s (24.3%) attending to their mock photometric images. We did not find remnants of later types than S0 in the sample taken from GalMer (i.e. there were not spirals). Therefore, we have a final sample of 96 S0-like relaxed remnants: 25 E/S0's and 42 S0s resulting from a major merger (67 in total), and 29 S0s after a satellite accretion. Nearly 67% of all S0-like remnants result from prograde orbits and 33% from retrograde encounters.

Table A.3 lists the visual morphological types assigned to the 202 relaxed remnants selected from the GalMer database. The artificial photometric images and particle-position maps for the face-on and edge-on views of the remnants finally classified as S0 and E/S0 galaxies can be found in Appendix B.

## 5. Other physical properties of the remnants

We also checked that the residual SF and gas contents of the selected S0-like remnants, as well as their global structure and kinematics, were consistent with those observed in nearby S0s. Although these properties will be discussed in detail in a forthcoming paper, we briefly comment them here.

As reported by previous merger simulations, the gas of the progenitors easily loses angular momentum during the encounter and falls towards the remnant centre, fuelling strong and brief central starbursts that produce new nuclear stellar components (e.g. Mihos & Hernquist 1994a,c,d, 1996; Hibbard & van Gorkom 1996; Barton et al. 2000; di Matteo et al. 2008; Tutukov et al. 2011; Vshivkov et al. 2011; Mapelli et al. 2015). At the end of the simulation, these new stellar sub-components have relatively young ages (<1–2 Gyr typically). This fact, together with the assumptions adopted to estimate the $M/L$ ratio in each particle, makes them govern the light emission of the remnant at the central regions (see some examples in Sect. 7.2). The accumulation of residual SF and small amounts of gas at the centres of ellipticals and S0s is not rare, and in fact becomes more common as the redshift increases (Trager et al. 2000; Yi et al. 2005; Fritz et al. 2009; Kannappan et al. 2009; Huertas-Company et al. 2010; Wei et al. 2010; den Heijer et al. 2015). We have checked that the levels and spatial distributions of residual gas and SF in our S0-like remnants are consistent with observations.

These young central components make our S0-like remnants brighter than real S0s with similar masses by ~1.5–1 mag in the

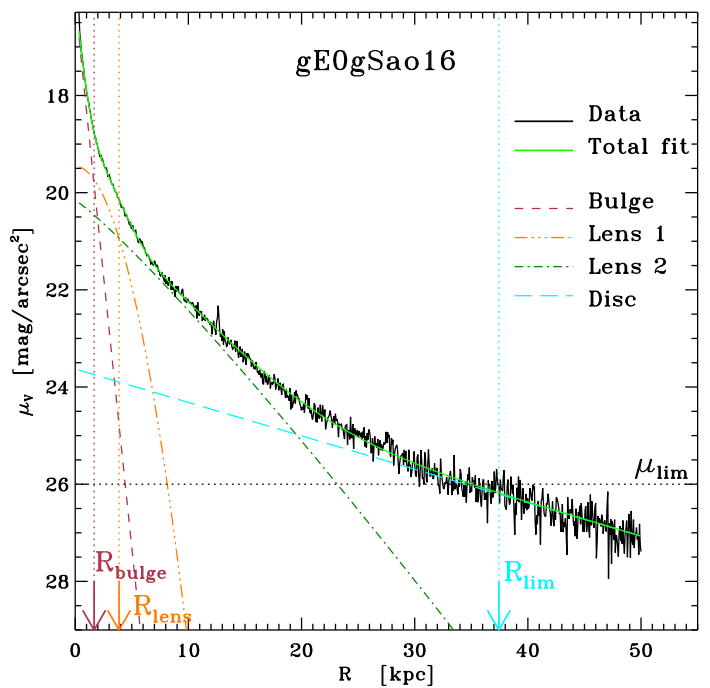

**Fig. 4:** Example of the multi-component decomposition performed to the 1D surface brightness profile of the remnant of model gE0gSao16 (E/S0) in the *V* band. The final galaxy can be well described using a Sérsic bulge, an outer exponential disc, and two additional intermediate Sérsic components (see the legend). The latter ones represent the contribution of a couple of lenses, which can be identified in its *V*-band images (see Fig. 5). The outer one seems to be spheroidal in the edge-on view in that figure, thus corresponding to a stellar halo. On the basis of this decomposition, we have defined the outer radii of the bulge, the innermost lens, and the outer disc limit of this galaxy ($R_{bulge}$, $R_{lens}$ and $R_{lim}$, respectively) as described in the text. The three outer radii have been marked with arrows in the Figure.

*R* and *K* bands, so their average $M/L$ ratios are twice lower than those of their real counterparts (B14; Q15a). However, similar $M/L$ values have been found in local S0s with residual central SF (den Heijer et al. 2015), so the remnants are not rare within their class. Moreover, the bulk of real S0s must have experienced longer relaxation times than the GalMer remnants, at least for ~1–2 Gyr more (see references in Sect. 1). Therefore, in case of allowing the models to evolve in isolation for an additional couple of giga-years, the young stellar populations at the core would quickly fade and redden enough to be completely embedded by the underlying old stellar population of the bulge (Prieto et al. 2013), making the magnitudes of our remnants more similar to those of their local analogues (as shown in T17).

The 96 S0-like remnants also present normal structures attending to their radial surface brightness profiles, as well as typical rotation curves and velocity dispersion profiles, with clear signs of hosting kinematically-decoupled components at the centre, which is also frequent in real S0s. For more information, see Eliche-Moral et al. (in prep.).

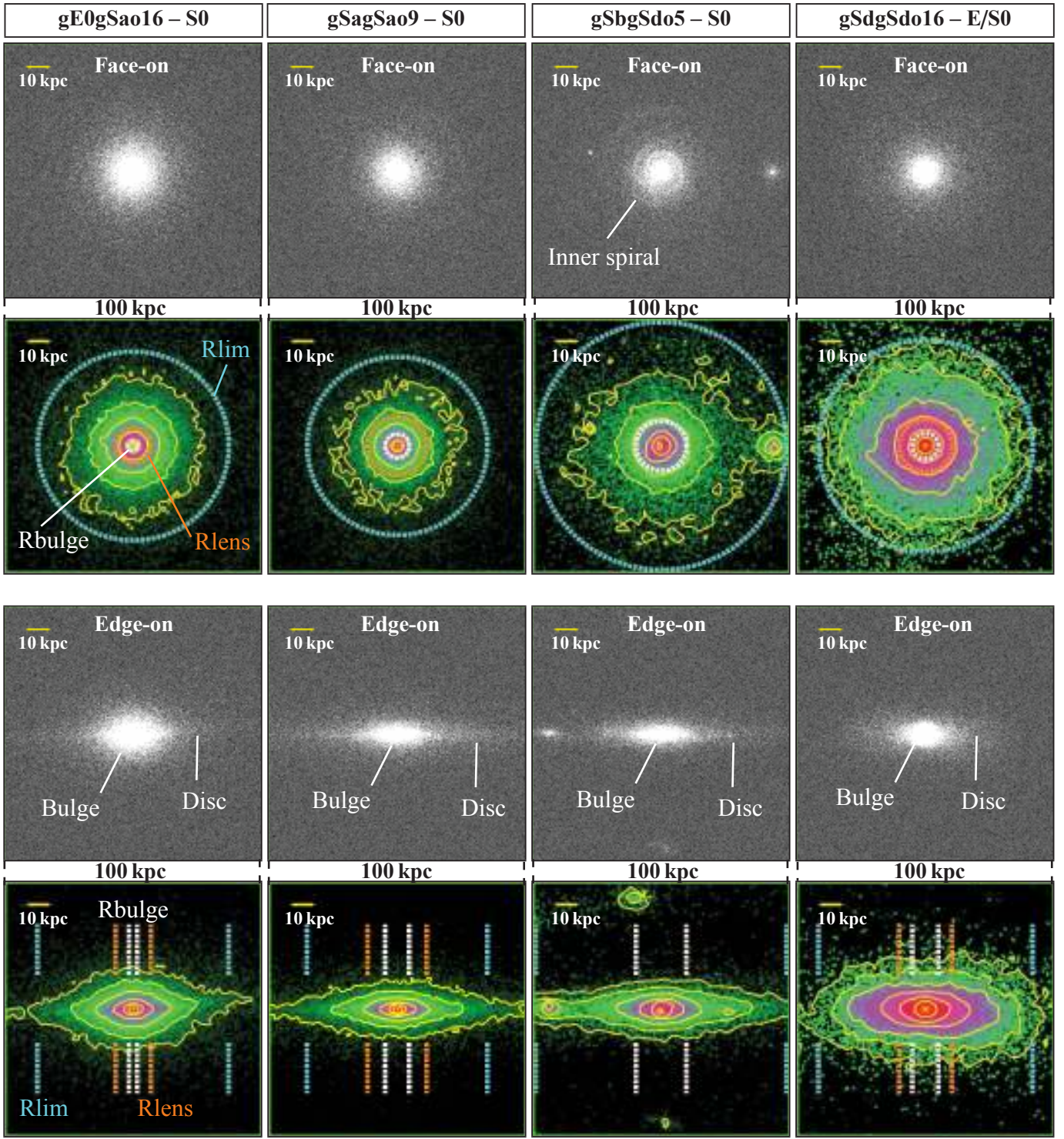

Fig. 5: S0-like remnants resulting from major mergers after ~3–3.5 Gyr of total evolution, showing global relaxed bulge+disc morphologies characteristic of real S0s and no traces of their past merger just ~1–2 Gyr after the full merger. Their face-on and edge-on $V$-band mock images are represented using a logarithmic grayscale that saturates the central regions to highlight the outskirts (*first and third rows of panels*). Below them, we have represented the same views of each remnant, but smoothing the images with a Gaussian with $\sigma = 10$ pixels (equivalent to 2.5") and using a logarithmic colour scale to emphasise the different brightness level of the different components (*second and fourth rows*). The location of $R_{bulge}$, $R_{lens}$, and $R_{lim}$ have been marked in the colour images in each case, using white, orange, and cyan thick dashed lines. Some isophotes have been over-plotted (*yellow*) to remark the real extent of the components in the original images, difficult to appreciate due to the grainy structure of the outer disc. The visual morphological types assigned to the remnants are indicated in the top labels. The FoV is 100×100 kpc$^2$ and $\mu_{lim}(V) = 26$ mag arcsec$^{-2}$ ($S/N = 5$) in all cases.

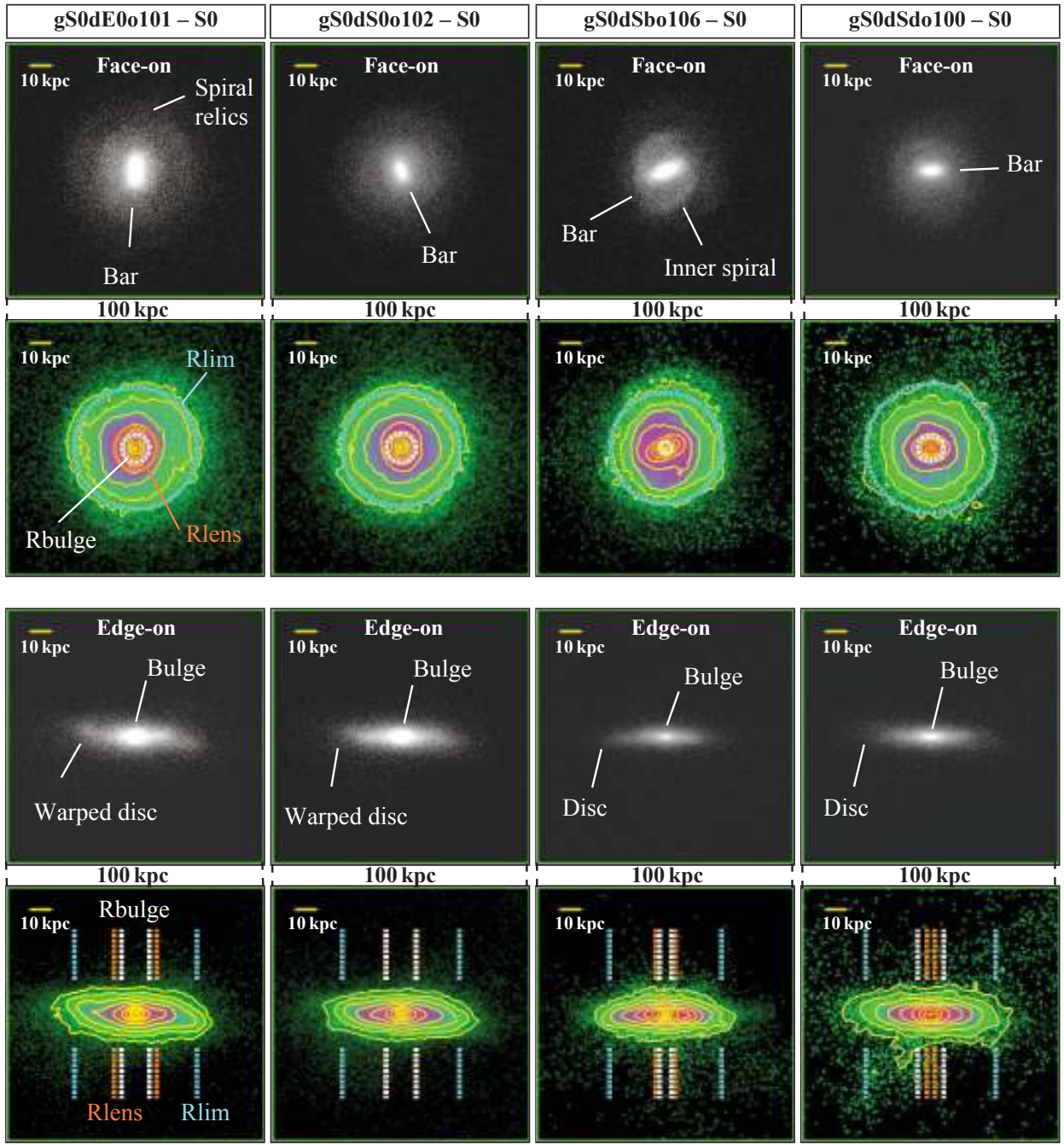

Fig. 6: *V*-band images of some S0 remnants resulting from minor mergers onto a gS0 galaxy, showing no traces of their past satellite accretion just after ~3 Gyr of total evolution. The bar in these remnants is an evolved version of the one already present in the gS0 progenitor. Warped discs in minor-merger cases are more common than in major mergers. Some of them can be seen in the edge-on views represented here (in particular, models gS0dE0o101 and gS0dS0o102). Notice the bending of the isophotes in their Gaussian-smoothed colour images. For a complete description, see the caption of Fig. 5.

## 6. Visual identification of morphological features

We visually inspected the morphologies of our S0-like remnants using the mock photometric images to identify particular morphological features.

Their global appearance, visual morphological components (including inner ones), and detectable tidal features have been compared with those observed in local S0 and E/S0 galaxies.

The 96 remnants were selected because they exhibited relaxed bulge+disc structures, with discs free of relevant spiral patterns and star forming knots (Sect. 4.2). So, we checked whether their bulge+disc structures were typical of S0 and E/S0 galaxies or not, and if they exhibited components other than the bulge and the disc, such as bars, lenses, ovals, stellar halos, rings, pseudo-rings, and inner spirals.

All these identifications were done visually using the face-on and edge-on $V$-band images, although the images in the other four broad bands ($U$, $B$, $R$, and $K$) were used to confirm them. Additionally, other diagnostics were used to check ambiguous cases, such as the multi-component decompositions performed to their azimuthally-averaged radial surface brightness profiles in $V$ and the rotation curves and velocity dispersion profiles of the final remnants (Eliche-Moral et al., in prep.).

As the majority of real S0s have a lens or oval (see Sect. 1), we paid special attention to this kind of sub-structures in our remnants. A lens or oval is a light subcomponent with a constant surface brightness surrounding or overlapping the bulge, which stands out from the exponentially-decaying profile of the external disc. Lenses are axisymmetric, whereas ovals present an elongation which usually coincides with the major axis of a bar (Laurikainen et al. 2005; Sil'chenko 2009, KB12). We have also identified rings and pseudo-rings in the images. We inspected the views in which the main disc of each remnant was edge on, looking for significant warps and flares in it.

We obviously identified bars too. Beside the large ones that extend through a significant part of the disc ($\sim$10–20 kpc), we also identified nuclear bars with sizes $\sim$1–2 kpc. However, the effects of the softening length of the simulations ($\epsilon = 200$ and 280 pc) make the properties of such small components quite uncertain. Concerning this, we have also found many nuclear compact sources in the remnants resulting from gas-rich mergers, with radii lower than $\sim$1 kpc, which produce a characteristic steep rise at the centre of the radial surface brightness profiles of the galaxies. However, any sub-structures with sizes lower than $\sim$1 kpc may be artefacts due to the softening of the simulations.

As commented in Sect. 5, some remnants develop small discs embedded in the bulge structure with sizes $\lesssim$ 5–6 kpc, as result of the central starbursts induced by the merger. Since they are made of young stars basically, they out-shine in some cases the central bulge and become visually detectable. We will refer to them as inner discs, but we stress that they are sub-components different from the inner exponential sections of the main discs in those remnants exhibiting Type-II, Type-III, or hybrid profiles (i.e. non-pure exponential –Type-I– profiles). For more information on Type-III profiles in our remnants, see B14.

Many of these inner discs contain disc-related phenomena, such as nuclear bars (as commented before) and inner spiral patterns, which can be visually detected. Sometimes, these features are in fact the signs pointing to the existence of an inner disc that is totally embedded in the bulge light distribution (i.e. not directly visible). We checked that all inner discs that could be visually detected in the images produced characteristic features in the kinematic maps of the remnants, such as $\sigma$ dips and peaks or decrements in the rotation velocity profiles, depending on whether the inner disc is co- or counter-rotating. However, the opposite is not always true (Eliche-Moral et al., in prep.). We exclusively focus on the inner discs that are directly detectable in the photometric images in the present study.

We have also surveyed the $V$-band images of the remnants to identify the tidal structures clearly pointing to their merger origin. However, due to our sample selection, any remnants with non-relaxed morphologies have been excluded from the sample, so none of them are at intermediate merger stages, but at very advanced ones (Sect. 3.1). This means that there are no remnants in our S0-like sample with strong tidal tails, double nuclei, huge shells, or "train-wreck" morphologies, although they can exhibit small tidal tails, plumes, and other weak signs of their past merger activity (i.e. "fine tidal structures", see Duc et al. 2015). In particular, we have looked for shells, tidal tails, disc asymmetries, plumes, tidal debris, and collisional rings. All these structures have been identified on images with $\mu_{\rm lim}(V) = 26$ mag arcsec$^{-2}$ (Vega) for a $S/N = 5$ (see Table 4), so they correspond to even fainter magnitudes in reality, since the human eye is extremely sensitive to faint structures (near the limit of $S/N \sim 1$). We have also identified the cases with tidal satellites within 100 kpc around.

In order to facilitate the identification of different morphological features in the $V$-band images of the remnants, we have represented in the figures throughout Sect. 7 the outer radii of three characteristic components of the majority of the S0-like remnants: the bulge, the outer disc, and the lens or oval that all galaxies seem to have. These characteristic radii have been defined according to the multi-component decompositions performed to the 1D surface brightness profiles in $V$ of the S0-like remnants.

First, the outer radius of the bulge ($R_{\rm bulge}$) is the radial position at which the fitted bulge component stops dominating the total surface brightness profile, compared to the fitted lens or oval contribution (in case there is one) or the fitted disc profile (in case there is not lens or oval or if it is completely embedded within the bulge light distribution). The outer lens or oval radius ($R_{\rm lens}$) was defined analogously, but considering the position at which the fitted brightness of the disc starts dominating over the fitted innermost lens or oval component. If the lens or oval was completely embedded in the bulge and disc light, it was not defined, even though it was necessary in the profile decomposition. In case of existence of several lenses or ovals, $R_{\rm lens}$ was defined for the innermost one. Finally, the outer disc radius ($R_{\rm lim}$) was defined as the radial position where the surface brightness of the fitted disc reaches the limiting surface brightness of the image (in our case, $\mu_{\rm lim}(V) = 26$ mag arcsec$^{-2}$, see Table 4).

Although these decompositions will be presented elsewhere, the one performed to model gE0gSao16 (S0) is shown in Fig. 4 as an example. In particular, this remnant required a couple of Sérsic components at the centre (besides a Sérsic bulge and an exponential disc) to be properly described. These two additional sub-components can be identified with the lenses observable in its face-on $V$-band images in Fig. 5. We have marked the locations of $R_{\rm bulge}$, $R_{\rm lens}$, and $R_{\rm lim}$ in Fig. 4. These characteristic outer radii delimit the transition between adjacent components (bulge–lens, lens–disc, disc–sky noise), so they are useful for their visual identification when over-plotted in the images, as shown in Sect. 7.

## 7. Results

The results are presented in Sects. 7.1–7.3. In Sect. 7.4, we also analyse the most relevant trends of the morphology with the initial conditions of the merger experiments. Throughout the text, we directly compare the morphology of our S0-like remnants in the $V$ band with the $K$-band images of real S0s in L11 and L13. We have checked that the global morphologies of our objects are very similar in both bands, a fact that allows this comparison (the *K*-band mock images of all S0-like remnants are available in Appendix B).

### 7.1. Global morphology: lack of bars in S0-like remnants from major mergers

In Figs. 5 and 6, we represent the face-on and edge-on *V*-band images of some S0-like remnants resulting from major and minor mergers, respectively, using a logarithmic grayscale that saturates the central regions to emphasise the fainter features at the outskirts of the remnants. This also allows a proper visualisation of their global bulge-disc structures (see also Appendix B). The grainy structure of the original *V*-band images in the outskirts make difficult to appreciate the real extent of the outer discs in the face-on views, which is better noticed in the edge-on images (Sect. 4.1.3). Therefore, we also show the result of performing a Gaussian smoothing with $\sigma = 10$ pixels (2.5") to each image in the same Figures, using a logarithmic colour scale that emphasises the fainter regions at the outskirts even in the face-on views. Some isophotes have been over-plotted in these smoothed images to stress the real extension of the components in the original images.

Both Figures demonstrate that the S0-like remnants have smooth and relaxed bulge+disc structures in general, very similar to those observed in local S0 galaxies (see, e.g. L11 and Buta et al. 2015). These simulations thus indicate that major mergers can produce remnants with typical present-day S0 or E/S0 morphologies only ~1–2 Gyr after the full merger. Although the original structures of the merging galaxies are basically destroyed during the encounter, a disc in the remnant is quickly rebuilt around the central spheroid (see Fig. 7 in Q15a). As commented before, none of the 202 relaxed remnants exhibits significant spiral patterns or SF knots spread in the disc, although some resulting from major encounters have faint spiral relics in the outer discs (see Figs. 5 and 6). This is also found in some real S0s, such as NGC 7371 (see Fig. 14 in L11).

Interestingly, none of the major-merger S0-like remnants develops a notorious bar at the end of the simulation either (Fig. 5), despite some of the encounters have significant gas fractions (in particular, the gSd+gSd ones). Their morphologies are similar to those of real unbarred S0s, such as NGC 4382, NGC 4546, NGC 4649, or NGC 5898 (see Fig. 5 in L11). Notice that we exclude small nuclear bars here, which are commented in Sect. 7.2. The only case with some relics of a large bar is gSbgSbo72 (see Fig. 7). It clearly shows an X-shaped structure in its edge-on images, but the bar is so diluted in the original images that it is hardly recognizable in the face-on views, except for the edges that stand out beyond the inner disc. These features are better seen in the Gaussian smoothed images depicted in the bottom panels of the Figure. This galaxy resembles NGC 2787 and NGC 1079 (see Fig. 11 in L13). Therefore, although major mergers are violent and many of them develop transient bars while the galaxies approach, the disc instabilities are rapidly diffused or inhibited in the remnant (Di Matteo et al. 2007). This may also be the reason after two additional facts: first, we have not found major-merger remnants with significantly flared discs, and secondly, very few of them exhibit significant warps in their discs (see Sect. 7.2.3).

The minor mergers onto the gS0 progenitor take longer to relax than major encounters in general (Sect. 3.2). In spite of this, we have found 29 cases of satellite accretions that are completely relaxed after only ~3 Gyr of total evolution. They all exhibit a strong bar, which is an evolved version of the one already present in the gS0 progenitor at the start of the simulation (see Fig. 6). Two faint spiral arms coming out from the bar edges are visible in many of their face-on images. The accreted dwarves induce transitory spirals and other bar-like instabilities in the discs at intermediate stages of the mergers (mainly during the pericentre passages), affecting the existing bars. Therefore, although minor mergers are weak interactions in the sense that they do not destroy the existing progenitor disc (see Fig. 7 in Q15a), they can significantly modify its dynamical structure (see Eliche-Moral et al. 2011, and references therein).

Figure 8 shows the number of remnants as a function of their visual morphological type in the sample of 202 relaxed remnants. Approximately half of this sample (including both major and minor mergers) has been classified as S0-like remnants (~35% as S0 and ~12% as E/S0, ~47% in total), while the other half are ellipticals. Since all remnants resulting from a gS0+dwarf merger produce a S0 remnant (as previously commented), the percentages are obviously less biased towards S0 types if we only consider the major-merger cases: ~39% of the 173 major-merger models have S0-like morphologies at the end of the simulations, the rest being ellipticals. It must be kept in mind that all relaxed remnants in our selection exhibited disc components in their projected density maps. In other words,

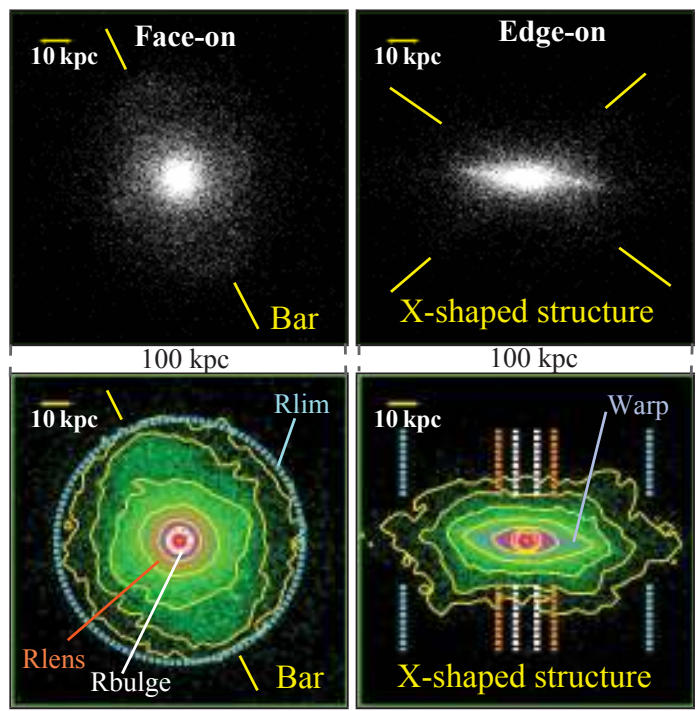

**Fig. 7:** *V*-band images of the unique major-merger remnant in our sample with clear relics of a large bar in its disc (model gSbgSbo72, S0 type). We have used a totally saturated logarithmic grayscale in the top panels to emphasise the faint structure of the outer disc. The bar forms a counter-clockwise angle of ~30° with respect to the vertical in the face-on image and produces an X-shaped structure in the edge-on view, typical of bars. These features are better appreciated in the Gaussian-smoothed images of the bottom panels ($\sigma = 10$ pixels or 2.5"), just attending to the shape of the over-plotted isophotes at the outer region. Both bar-related features have been marked with yellow guidelines in the corresponding panels. The FoV is $100\times100$ kpc$^2$ and $\mu_{\rm lim}(V) = 26$ mag arcsec$^{-2}$ ($S/N = 5$). For more information, see the caption of Fig. 5.

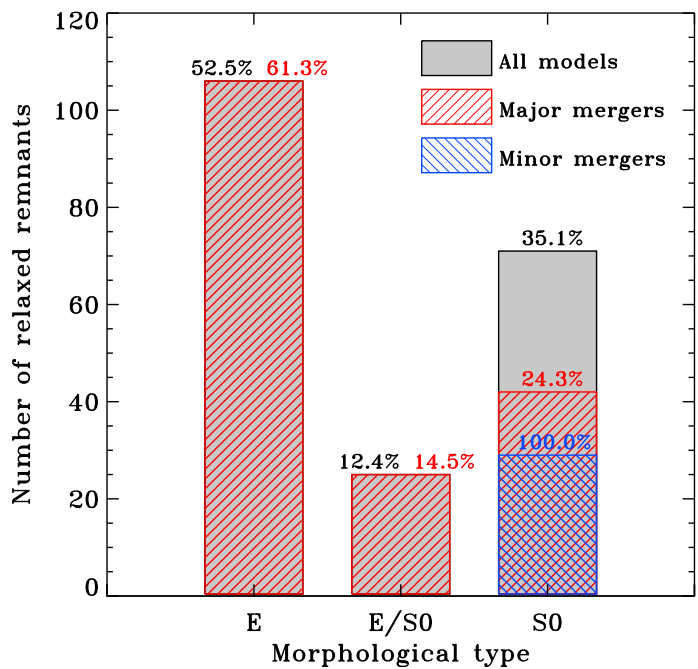

Fig. 8: Statistics of the 202 relaxed remnants as a function of their visual morphological types, for the total, major-merger, and minor-merger selected samples separately (see the legend in the panel). Percentages are computed with respect to the total number of models in each sample.

there are more relaxed elliptical remnants in the database which are not included in our sample.

In conclusion, the remnants of major mergers (< 3:1) can experience such efficient relaxation that they can morphologically resemble local S0s just ~1–2 Gyr after the full merger. Additionally, we have found that none of the S0-like remnants resulting from major mergers is significantly barred, despite the significant gas amounts present in some cases. These results are discussed in Sects. 9.1 and 9.4.

### 7.2. Morphological features and ICs of the S0-like remnants

In Fig. 9, we represent the face-on photometric images in $V$ of some S0-like remnants, using different colour scales and FoVs to highlight different sub-structures found in them. The ICs which are visible in the views shown in the first and third rows of panels of the Figure (in grayscale) have been labelled. The views in the second and fourth rows are analogous in FoV and line of sight to those represented in the first and third rows, respectively, but applying a Gaussian smoothing to the images, using a colour scale and over-plotting isophotes to emphasise the real extent of the sub-structures labelled in the corresponding grayscaled panels. We have also indicated the location of $R_{bulge}$, $R_{lens}$, and $R_{lim}$ in these views to help the identification of the sub-components. The bottom row represents a zoom into the innermost $10 \times 10$ kpc$^2$ of each remnant, using a staircase colour scale that emphasises the structure at their cores. All ICs visually detected in the S0-like remnants are listed in Table A.4 (see Appendix B too).

We have found that the S0-like remnants of both major and minor mergers exhibit complex central structures and host a wide diversity of ICs detectable in their photometric images, such as lenses, ovals, inner discs, inner spirals, rings, pseudo-rings, embedded inner discs (inclined or not), nuclear bars, and compact sources, as it is often observed in real S0s (Scorza et al. 1998; Scorza & van den Bosch 1998; Afanasiev & Sil'chenko 2000; Rest et al. 2001; Erwin & Sparke 2002; Erwin et al. 2003; Sil'chenko 2002; Sil'chenko et al. 2002; Sil'chenko 2015; Erwin & Sparke 2003; Emsellem et al. 2004; Falcón-Barroso et al. 2004; Kormendy & Kennicutt 2004; Balcells et al. 2007; Sil'Chenko et al. 2011; L09; L10; L11; L13). Concerning the nuclear bars and compact sources at the centres, we remark that they could be affected by the effects of the softening of the simulations, as commented in Sect. 6.

We have computed the percentages of the different morphological components visually identified in the remnants for the major- and minor-merger samples separately. The results are listed in Table A.5. More than 58% of the major-merger S0-like remnants host ICs that can be detected visually in the images, whereas only 3% of the minor-merger ones do so. Probably, the large bar and the associated oval present in all minor-merger remnants are hiding any additional ICs in them. We comment the results in the Table extensively in the next Sections.

Moreover, although the majority of the lenses in the major-merger cases look like embedded discs when viewed face-on, their vertical structures differ in the edge-on images, as shown in Fig. 11. In ~21% of the S0-like major-merger remnants, the lenses correspond to a spheroidal stellar halo surrounding the bulge. However, ~64% of them resemble vertically-thin (flat) disc-like structures instead. The remaining major-merger cases are intermediate between these two extremes: they are lentil-shaped and vertically thick. However, all ovals in the minor-merger S0 remnants are vertically thin. This diversity of vertical structures of the lenses is consistent with the properties of the lenses observed in real S0s too (Erwin et al. 2005).

The percentages of lenses and ovals according to their kinds of vertical structures are represented in panel (a) of Fig. 10 for the major- and minor-merger samples. The kind of lens or oval (spheroidal, flat, lentil-shaped) of each model is indicated in Table A.4.

#### 7.2.1. Lenses and ovals

Besides the bulge and the disc, all S0-like remnants resulting from major mergers exhibit a lens or oval component, despite lacking any strong bars. Some examples are provided in Fig. 9. They extend up to ~20–50% of the visual radius of the disc, exhibiting a sharp end, and are mostly axisymmetric. Only in two cases (models gE0gSao1 and gSbgSbo16), they look slightly elongated, thus being ovals instead (one of these cases is shown in the Figure). Unbarred S0s hosting lenses or ovals as those generated by these simulations are common in the local Universe (see, e.g. NGC 524, NGC 2880, NGC 4472, NGC 4552 or NGC 7192 in Figs. 5, 11 and 15 in L11).

Therefore, all S0-like remnants in our major-merger sample host a lens (~97%) or an oval (~3%). These percentages are represented in panel (a) of Fig. 10. Furthermore, the 29 S0 remnants resulting from minor mergers present an evolved version of the original bar of the gS0 progenitor, as commented before. These bars are surrounded by an oval component in all cases, as seen in Figs. 6 and 9. These results agree pretty well with the properties of real S0s, because ~97% of local S0s host a lens or oval component (L09; L11).

The flat lenses in the major-merger remnants seem associated with embedded inner discs in many cases (see Figs. 11 and 12).

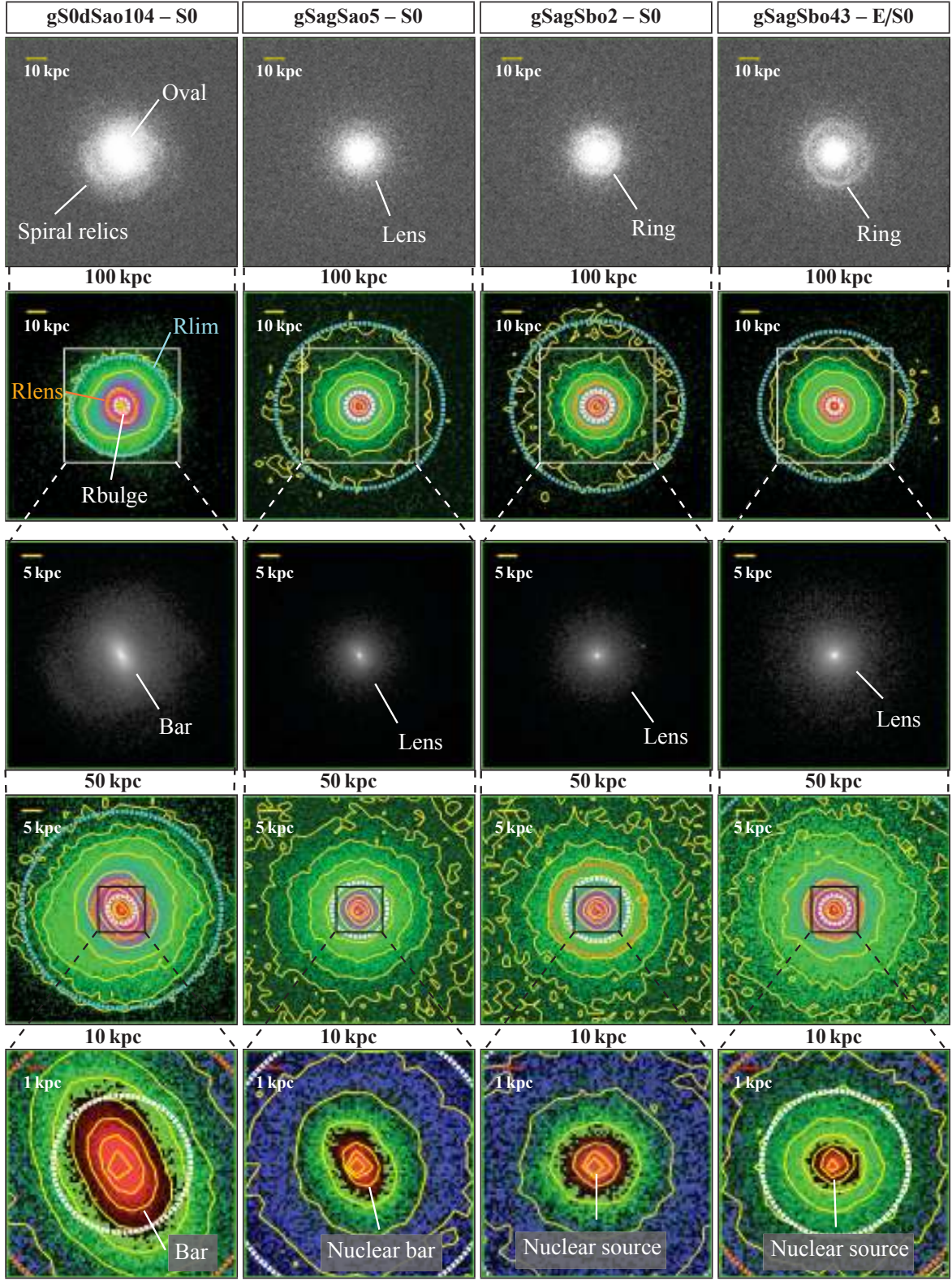

Fig. 9: Face-on *V*-band images of some S0-like remnants with observable ICs. Different zoom levels and colour scales have been used in the panels to emphasise ICs such as lenses, ovals, inner discs, inner spirals, inner rings, pseudo-rings, and nuclear bars. In all cases, $\mu_{\mathrm{lim}}(V) = 26\,\mathrm{mag\,arcsec}^{-2}$ ($S/N = 5$). *First row*: original images (with no smoothing) for a FoV of $100{\times}100\,\mathrm{kpc}^2$, using a logarithmic grayscale that saturates at the centre to highlight the structures at the intermediate galaxy body, such as inner spirals, pseudo-rings, and inner rings. Some lenses and ovals are also noticeable in these views. *Second row*: same images as in the top panels, but using a Gaussian smoothing with $\sigma = 10$ pixels (or 2.5") to emphasise the faint outer structures. The logarithmic colour scale used here, the isophotes (*yellow*), and the locations of $R_{\mathrm{bulge}}$, $R_{\mathrm{lens}}$, and $R_{\mathrm{lim}}$ delimit the real extension of the structures in the top panels. [**Note.–** *Caption continues.*]

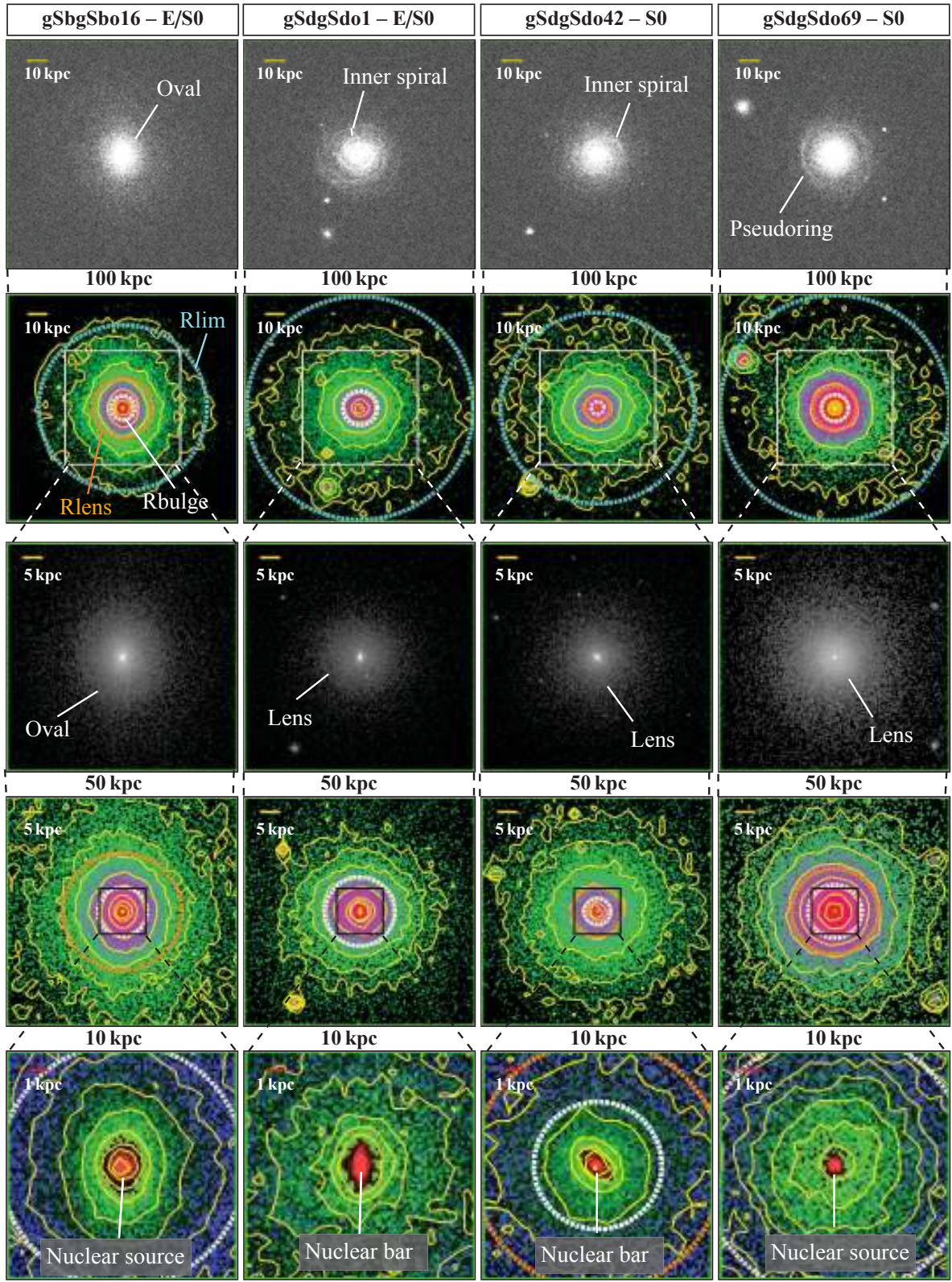

Fig. 9: [Cont.] [*Caption continuation.*] *Third row*: magnification of the central $50\times50\,\mathrm{kpc}^2$, applying a Gaussian smoothing of $\sigma = 4$ pixels (equivalent to 1") and using a logarithmic grayscale that highlights the ovals and lenses (the outer disc is masked by the lowest level). All major-merger S0-like remnants have an axisymmetric lens, except for two, which host an oval instead (one is shown here: gSbgSbo16). All minor-merger S0 remnants have an oval surrounding the central bar, as gS0dSao104. *Fourth row*: same images as in the third row, but using a Gaussian smoothing with $\sigma = 4$ pixels (1") and a colour scale. The isophotes and the outer radii help to identify the lenses and ovals of the third row. *Fifth row*: magnification of the innermost $10\times10\,\mathrm{kpc}^2$ of the original images. The logarithmic staircase colour scale unveils the structure of the galaxy core, revealing nuclear bars with sizes of ∼1–2 kpc and nuclear compact sources (in red, possibly affected by the softening of the simulations), as well as embedded inner discs (there are faint inner spiral patterns in gSdgSdo1 and gSdgSdo42, in dark green).

It is not clear just through visual inspection of the images whether the lenses and inner discs are independent substructures or not, in other words, whether the flat lenses are outer stellar envelopes containing the inner discs or just part of the same component. Although their light distributions are difficult to separate in many radial surface brightness profiles, we have found some cases that allow independent fits for the lens and the embedded inner disc, pointing to the fact that they may be relatively independent (see Eliche-Moral et al., in prep.).

In conclusion, lenses and ovals are ubiquitous in these simulations, in both major and minor mergers. These components are usually considered as relics of bar evolution, but these models demonstrate that they can result from major mergers too, without requiring the presence of any bars. We discuss this result in Sect. 9.3.

### 7.2.2. Embedded inner discs and related features

Some inner discs can be directly observed in the mock images of the remnants (as those represented in Fig. 12), while others cannot because they are completely buried in the bulge light. We know they exist because all these dynamically-cold inner structures produce clear traces at the centre of the kinematic profiles of the remnants that reveal their presence, such as $\sigma$ dips or counter-rotation (see Sect. 6). However, we exclusively focus on those that can be directly detected in the images here, which are listed in Table A.4.

The majority of these inner discs seem co-planar with the main discs. Nearly 24% of the S0-like major-merger remnants have embedded inner discs within the main galactic plane that are visually detectable. Moreover, an additional 21% of the major-merger S0-like remnants host inclined inner discs that can be observed in the images. Some examples of both kinds of inner discs are represented in Fig. 12. Their physical radii are ≲2–3 kpc. These models with edge-on inclined inner discs resemble some real cases, such as NGC 1161, NGC 4638, NGC 5493, and NGC 7029 (Fig. 5 of L11). The inclined inner discs in our sample always result from highly-inclined ($i > 45°$) retrograde orbits, except for three cases: one with a co-planar retrograde orbit (gSdgSdo51) and two highly-inclined encounters, but in prograde orbits (gSbgSbo22 and gSdgSdo21).

Many of these inner discs contain disc-related phenomena, such as faint inner spirals, inner rings, pseudo-rings, or nuclear bars, as those seen in Fig. 9. The inner spirals are very low-level structures that may extend up to ∼ $0.5R_{25}(V)$ (∼10–15 kpc, depending on the model). The inner rings and pseudo-rings tend to appear at the limiting region between two galaxy components, such as a lens–lens, lens–disc, or disc–disc transitions (in case of being non-pure exponential disc), as observed in many real cases (Laine et al. 2014). The nuclear bars in these simulations have sizes <1–2 kpc, so they are probably affected by the softening of the simulations (Sect. 6). The morphological features identified in each remnant are listed in Table A.4. Similar structures are found in the centre of many unbarred S0s, such as NGC 1543, NGC 2782, NGC 4220, NGC 5953, NGC 7213, and NGC 7742 (Figs. 5, 13, 14 and 16 in L11).

We have found inner spirals in 18% of the S0-like remnants from major mergers, inner rings in 3%, inner pseudo-rings in 7%, and a total of 13% with nuclear bars. Concerning the minor mergers, we have found only one pseudo-ring (3%) among their remnants. As commented before, the large bar+oval structure in these models may be obstructing the visual detection of any embedded ICs. The percentages of S0-like remnants hosting inner discs and related phenomena (inner spirals, inner rings, nuclear bars, etc.) are compared for the major- and minor-merger samples in Fig. 10 (panel c).

Therefore, the remnants present complex central structures, hosting several ICs such as circumnuclear discs, inner rings and pseudo-rings, inner spiral patterns, nuclear bars, and compact sources. Previous numerical simulations have already proven that minor mergers onto S0 galaxies can produce all kinds of ICs (e.g. Eliche-Moral et al. 2011; Mapelli et al. 2015), but the selected `GalMer` experiments specifically demonstrate that these ICs can easily result from major mergers as well. This is discussed in Sect. 9.3.

### 7.2.3. Main disc features

In Fig. 10, we also compare the percentages of several large features present in the main disc of the remnants in the major- and minor-merger samples (panel b). In general, the main discs of the S0 remnants resulting from minor mergers seem to be more sensitive to instabilities than those resulting from major ones. As indicated in Sect. 7.1, the majority of the remnants in the minor-merger sample have warped main discs (68%), whereas only 4% of the S0-like remnants coming from major mergers exhibit them. Some minor-merger cases can be seen in the edge-on images of Fig. 6. Figure 13 represents two of the only three cases of major-merger S0-like remnants with warped discs (the other one is in Fig. 7).

The same panel in Fig. 10 shows that most minor-merger remnants present faint spiral signatures in the outer disc (∼79%), coming out from the bar edges in most cases (see Fig. 6), whereas we have found them only in 7% of the major-merger sample (Fig. 5). This is probably related to the lesser strict condition that we had to impose to minor mergers to be considered in dynamical equilibrium (see Sect. 3.3), which implies lower relaxing periods in minor mergers than in major ones (typically ∼1 Gyr vs. ∼2 Gyr, see Table A.3). Nevertheless, there are also major-merger cases with relaxing times of just ∼1 Gyr which do not exhibit warps or outer spiral relics either. For similar relaxing time periods, relaxation is more efficient in major encounters than in satellite accretions, and this obviously affects to the level of the instabilities in the remnant disc.

This difference in the stability of the remnant discs could be partially due to the different numerical resolution of the experiments, higher in minor mergers than in major ones (Sect. 2). However, the improvement of the numerical sampling of the disc in the former ones is only by a factor of ×2, so it may not be determinant, at least considering that the number of particles used in the major mergers is already high enough to reproduce pretty well other large-scale structures, such as the Type-III discs of real S0s (see B14). Moreover, previous SPH simulations of major mergers with gas, but without SF effects implemented, indicate that the trend of the gas is to settle into warped discs (Barnes 2002). In `GalMer` models, the merger-induced starbursts accumulate the gas at the central regions instead of dispersing it across the disc, preserving the stability of the (mostly stellar) remnant disc. Therefore, dissipation effects seem to contribute to the disc stability.

We can thus conclude that, in general, the remnant discs resulting from major-merger experiments are more resistant to disc instabilities than those resulting from minor mergers for similar relaxing times because of the strong dissipation undergone during the merger. We discuss this in Sect. 9.2.

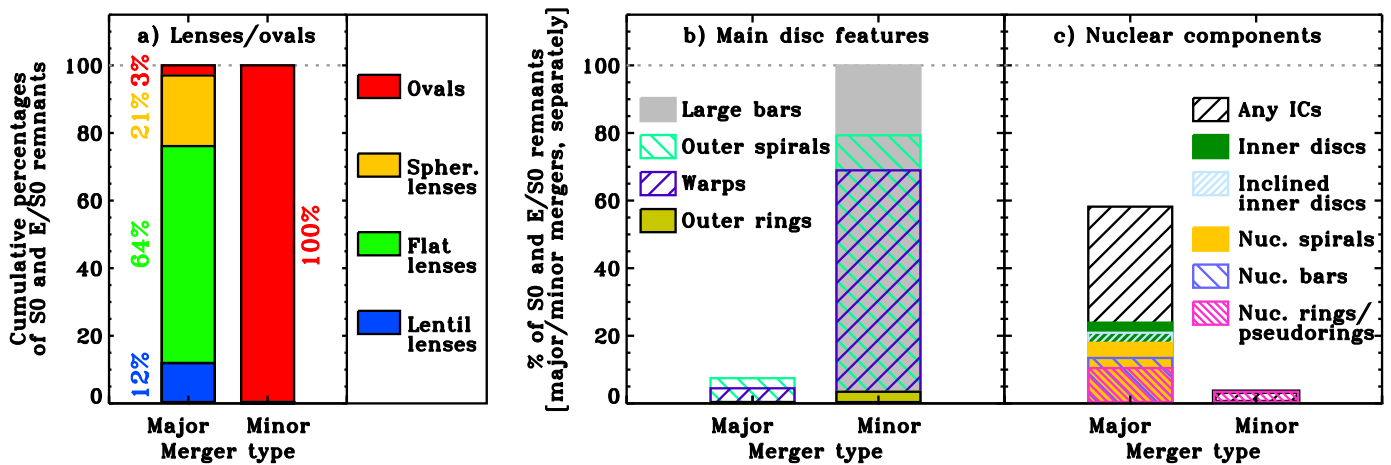

Fig. 10: Percentages of S0-like remnants resulting from major and minor mergers hosting different visual ICs at their centres or certain morphological features in their main discs. *Panel a*: cumulative percentages of lenses an ovals in the major- and minor-merger samples separately. We have distinguished the three kind of lenses according to their vertical shapes, as identified in their edge-on images (spheroidal, lentil-shaped, or flat). All S0-like remnants host a lens or oval component. *Panel b*: percentages of morphological features detected in the main disc of the S0-like remnants for the major- and minor-merger samples separately. This indicates that major-merger remnants tend to be more resistant to disc instabilities than minor-merger ones. *Panel c*: the same as in panel (b), but for the ICs. Major-merger S0-like remnants also tend to host a higher number of visual ICs than minor-merger ones. See the corresponding legend in each panel. More information in Table A.5.

### 7.3. Morphological traces of past merger activity

#### 7.3.1. Global statistics of merger relics

The morphological features of the S0-like remnants described in Sect. 7.2 (lenses, ovals, inner discs, warps, etc.) obviously derive from the merger event, but there are different evolutionary processes that can produce them in reality. On the other hand, other features are exclusively generated through interactions and mergers, such as tidal tails, shells, collisional rings, and severe disc distortions, thus pointing directly to a past merger event in any galaxy that exhibits them.

Some examples of S0-like remnants with merger relics visually detectable in the images are shown in Fig. 14. The asymmetric discs, plumes, ripples, small tidal tails, collisional rings, and tidal debris observed in the models are very similar to those exhibited by some real nearby S0s (see, e.g. NGC 2782, NGC 2902, NGC 3081, NGC 4293, NGC 4382, or NGC 7585 in Figs. 5–7, 11, and 15 by L11). In Table A.4, we have listed all visual morphological characteristics that can be directly attributed to the past merger experienced by each S0-like remnant.

We have computed the percentages of several merger-related features, both for the major- and minor-merger samples, as well as the total fractions of remnants with and without any merger-related signs in each sample (see Table A.5). These data are plotted in Fig. 15 (panel a). This Figure shows that the number of merger features detected in the major-merger remnants is very low: only ∼24% of the S0 and E/S0 remnants coming from a major merger present at least one morphological feature pointing to their past merger event, a surprisingly low fraction under the popular premise that these events are catastrophic and imprint long-lasting morphological distortions to the involved galaxies. This means that most remnants in the major-merger sample (∼76%) do not exhibit any traces of the merger that originated them just ∼1–2 Gyr after the full merger in relatively deep optical broad-band images. The dynamical relaxation is so efficient in them that the strong tidal features generated during intermediate stages of the encounter are quickly diluted after the full merger.

Conversely, ∼66% of the minor-merger remnants present at least one morphological trace pointing to their past merger activity in the photometric data (panel (a) in Fig. 15). Therefore, only ∼34% of them do not exhibit any traces of the accretion they have undergone in the images for identical photometric conditions as the major-merger ones. As commented in Sect. 3.3, the tidal interaction and dynamical relaxation in satellite accretions take more time than in major events, so their effects are detectable for longer time periods. Although the minor-merger remnants in our sample have lower relaxing time periods than those from major mergers on average (see Table A.3), the trend of major-merger remnants to lack of detectable tidal relics in the images remains even if we only consider those models with relaxing periods similar to those of minor mergers (∼1 Gyr).

Therefore, most S0-like remnants coming from major events do not show any morphological features revealing their past merger origin just ∼1–2 Gyr after the full merger, not even faint ones, at least for conditions typical of deep ground-based data ($\mu_{\mathrm{lim}}(V) = 26\,\mathrm{mag\,arcsec}^{-2}$ for $S/N = 5$, see Table 4) and distances as nearby as only 30 Mpc. Consequently, the observational fact that very few local S0s present significant merger relics in real data with similar depths (≲4%, see L11) does not necessarily imply that these galaxies cannot have derived from a major merger, at least if the encounter took place ≳2–3 Gyr ago. In fact, many of these local S0s may exhibit some tidal traces in much deeper data, as found in many cases (see, e.g. Duc et al. 2011, 2015, 2018). This means that the lack of noticeable merger traces in a present-day S0 (even for very deep photometric conditions) does not exclude a possible major-merger origin of the galaxy at all, because the merger could have occurred such long time ago that most of its relics could have been diluted by the present or be only detectable in prohibitively deep data.

However, the tidal features induced by minor mergers are longer-lasting than those from major events for similar relaxing time periods (∼1 Gyr) and observing conditions, despite being less destructive events. Therefore, minor-merger relics are more easily detectable than major-merger ones after long enough time periods for similar observing conditions. This means that the higher fraction of local S0s exhibiting minor-merger traces than major-merger ones (see references in Sect. 1) does not necessarily imply that minor encounters have been more relevant than major ones in the buildup of this galaxy population either. Minor events may have been more numerous than major ones at later epochs. In fact, the persistence of their relics would explain why they are detected more often in local S0s. But this does not exclude the possibility that many of these S0s were formed earlier and mainly through major mergers that have not left any visual morphological imprints in them at the present (at least, for typical observing conditions). We discuss these results in Sect. 9.2.

### 7.3.2. Tidal tails, merger debris, and tidal satellites

The tidal tails detected in the S0-like remnants resulting from both major and minor mergers are faint and small compared to the galaxy body in most cases (some examples are shown in Fig. 14). The most noticeable ones are presented by two minor-merger models (gS0dE0o105 and gS0dE0o106) and a major-merger one (gSagSdo2, see the Figure). Some of them contain a tidal satellite at the end of the tail (as model gSagSdo2). But, in general, tidal tails are faint and scarce in the sample: less than 10% of both major- and minor-merger samples exhibit them (see panel (b) in Fig. 15). Moreover, we have found only one remnant with a collisional ring and another with weak shells (the former is shown in Fig. 14).

We have also looked for tidal debris, that is widespread stellar material around the galaxy, expelled from it during the merger, which does not form a clear tail (Feldmann et al. 2008). Some examples are represented in Fig. 14. We have again found this feature more frequently in minor-merger remnants than in major ones (34% vs. 15%). This trend is followed by other merger-related features: plumes and small ripples are found in only 6% of the major-merger remnants, whereas they are present in 62% of the minor ones; only 1% of the major mergers present a disturbed disc, whereas 21% of the minor ones exhibit significant asymmetries in their discs (see panel (b) in Fig. 15).

We have not considered inclined discs, warps and tidal satellites as merger-related features in Tables A.4 and A.5, although they are known to result from mergers. Inclined discs can also be formed through external gas accretion with a high angular momentum (Sil'chenko & Afanasiev 2004), while warps can be produced by simple flybys or accretion of misaligned cold gas (Sánchez-Salcedo 2006; Mapelli et al. 2008; Kim et al. 2014; Gómez et al. 2016, 2017). Therefore, they cannot be considered inherent to mergers.

On the contrary, tidal satellites are intrinsically linked to merger and interaction activity, mostly if there is a high gas reservoir in galaxies (Bournaud & Duc 2006; Bournaud et al. 2007a, 2008b; Pawlowski et al. 2011). However, it is quite difficult to identify the tidal nature of a satellite orbiting a real galaxy just visually, because the tidal structures containing these satellites quickly fade in advanced stages of the merger (Hammer et al. 2010, 2013; Kaviraj et al. 2012). The tidal nature of these satellites is obvious in these simulations, but an observer could not have inferred it just attending to their photometric images: they look like independent orbiting satellites (see, e.g. Figs. 5, 9 or 13).

Nearly half of the S0-like major-merger remnants exhibit one, several, or a complete bunch of tidal satellites (see panel (b) in Fig. 15). They tend to appear in gas-rich models as ex-pected (see the previous references), although their number also depends on the orbital parameters of the encounter. On the contrary, no minor-merger remnant develops tidal satellites, because the gS0 progenitor is gas free and there is not enough gas in the satellites to produce the long gas-rich tidal tails that collapse into the star-forming clumps where these satellites form (see the same panel).

The only two S0-like remnants where some tidal structures remain around the satellites are models gSagSdo2 and gSdgSdo74 (see Fig. 14). The dramatic dilution of faint tidal tails and even of tidal satellites in the remnants when observing conditions are accounted for is illustrated in Fig. 16 for a couple of cases. This again points out the relevance of a proper simulation of the effects and limitations of real observational data in simulations to establish a fair comparison between the morphologies of real and simulated galaxies.

### 7.4. Trends of the final morphology with the initial conditions

#### 7.4.1. Initial gas fractions

Hopkins and collaborators reported that the survival of a disc in a major merger is a complex function of the initial conditions of the encounter, although it strongly depends on the gas content of the progenitors (Hopkins et al. 2009a,b). Figure 17 represents the number of S0 and E/S0 remnants found in the sample of 202 relaxed remnants as a function of the morphological types of their progenitors, for major and minor mergers separately. The models that have lower initial gas fractions are located on the left of the abscissas axis, their available gas content increasing towards the right.

The Figure indicates that the total number of relaxed S0-like remnants coming from major mergers increases towards gas-rich experiments in general, as expected. As numerous numerical studies have demonstrated, the gas acquires part of the orbital angular momentum of the encounter and tends to be deposited in a rotating disc at the centre. This explains why gas-rich encounters produce remnants with discs more frequently than gas-poor ones, which tend to result in ellipticals instead (see Mihos & Hernquist 1994d, 1996; Jesseit et al. 2005, 2007; Naab et al. 2006, among others). Therefore, although these simulations demonstrate that major mergers between progenitors of any types can produce a S0-like remnant, the likelihood increases if they involve high gas contents, consistently with previous studies.

On the contrary, the number of S0 remnants decreases towards experiments with more gas-rich satellites when minor mergers are considered (see Fig. 17). This is a selection effect: we looked for dynamically relaxed remnants. In minor mergers, the accretion is faster in more concentrated satellites for similar masses and orbital parameters (Aguerri et al. 2001; Eliche-Moral et al. 2006). Consequently, the accretion of an early-type satellite takes less time than a late-type one in these simulations, because the former is denser. The last case also requires longer time periods for relaxing, making it difficult to find a minor-merger experiment with a dSb or dSd that results in a relaxed remnant after ~3 Gyr in the `GalMer` database.

#### 7.4.2. Initial orbital parameters

In Fig. 18, we plot the fractions of S0, E/S0, and elliptical galaxies in the sample of 202 relaxed remnants as a function of the initial orbital parameters of the encounters. The top left panel of the Figure represents the trends with the orbital inclination

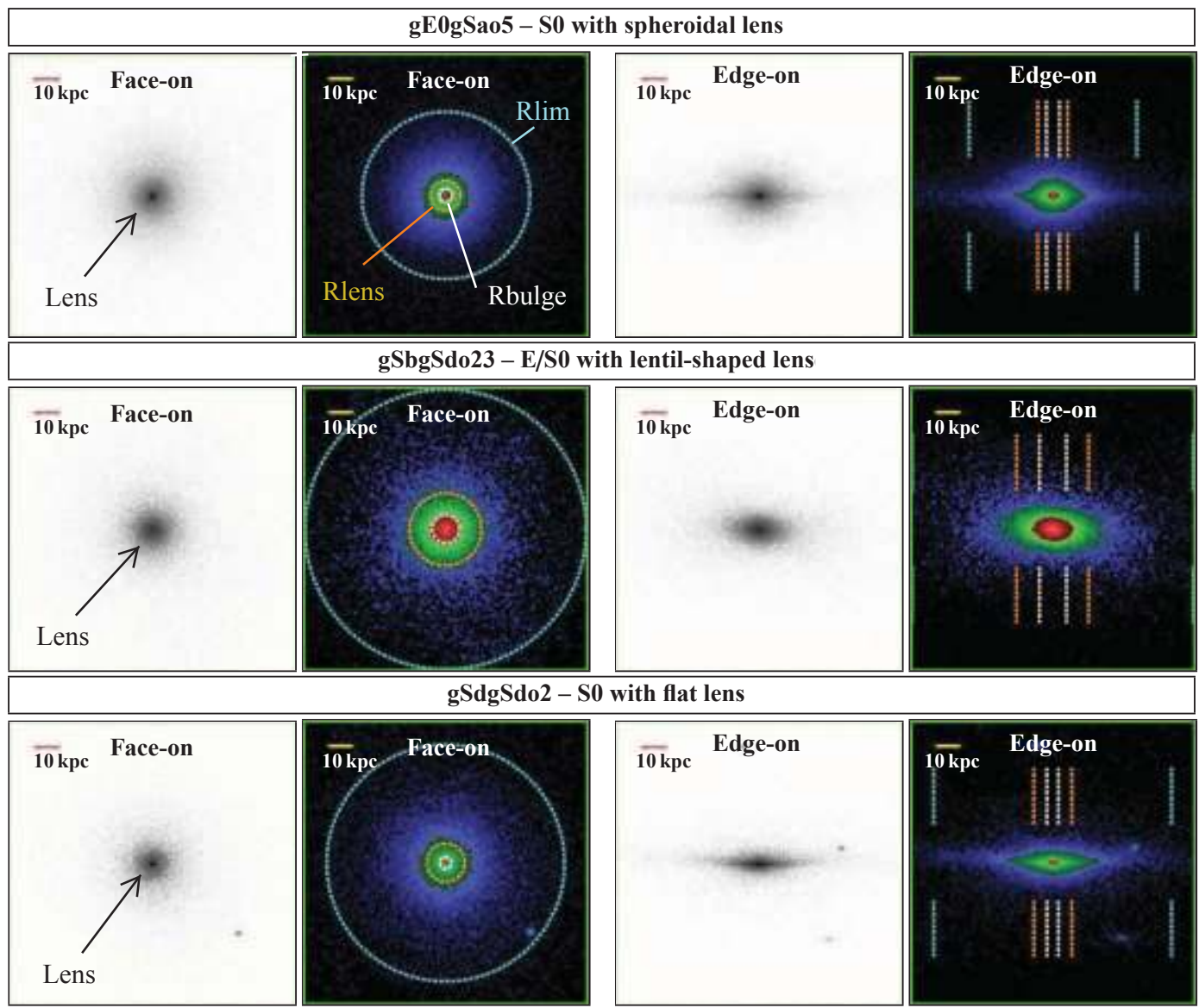

**Fig. 11:** *V*-band images showing the face-on and edge-on structure of the lens components in some S0 remnants of our sample. We represent each view using two different logarithmic colour scales: a grayscale to reveal the real appearance of the lenses in the *V*-band images, and a staircase colour scale in which the green level has been chosen to highlight the lens structure and its extension in each case. The frames with the colour scale have also been Gaussian smoothed assuming $\sigma = 10$ pixels (2.5"). The location of $R_{\rm bulge}$, $R_{\rm lens}$, and $R_{\rm lim}$ have also been indicated in these views. Model gE0gSao5 hosts an spheroidal lens (notice its extended, roundish structure in the edge-on view). The lens in model gSdgSdo2 is vertically flat, and it is probably related to the embedded inner disc that can be observed in the edge-on image. The lens in model gSbgSdo23 is between these two extremes (lentil-shaped). For more details on the images, see the caption of Fig. 5.

(*i*). The total fraction of relaxed remnants resulting from major mergers distribute quite similarly among the four inclination values used in this sort of experiments (0, 45, 75, and 90º), so the trends observed in each morphological type are not going to be biased by differences in the number of experiments carried out for each *i* value. We find that the percentage of S0 remnants resulting from co-planar encounters is much higher than for other inclinations, whereas the fraction of elliptical remnants increases with *i*. These trends were expected, since the likelihood of producing a remnant with a significant disc component in a merger is known to be higher when the progenitor discs are nearly co-planar with the orbit (Bournaud et al. 2004, 2005b). The fraction of E/S0 galaxies is quite similar for all inclinations. For minor mergers, the trend with *i* is mostly due to the lower number of relaxed models found for $i = 66º$ than for $i = 33º$, the only two values surveyed in these sort of experiments. Therefore, the main result we derive is that a S0 or E/S0 remnant can be formed from a major merger with any orbital inclination value, despite the trend to result preferentially from low-inclination orbits (otherwise, expected).

The top right panel of Fig. 18 represents the fraction of relaxed remnants as a function of the initial kinematic energy of the encounter, for each morphological type. The total fraction of relaxed remnants noticeably decreases when this energy increases in major mergers, consistently with previous simulation studies. For higher values of the initial relative velocity between the progenitor galaxies, longer periods of time are required for getting a fully merged and relaxed remnant. Consequently, many experiments with high values of this parameter are still merging or flying by at the end of simulations. Nevertheless, the fraction of relaxed remnants in minor mergers seems to be independent of the initial kinematic energy of the encounter, probably because the lower inertia of the dwarfs facilitates the accretion by the progenitor gS0, independently of their relative initial velocity.

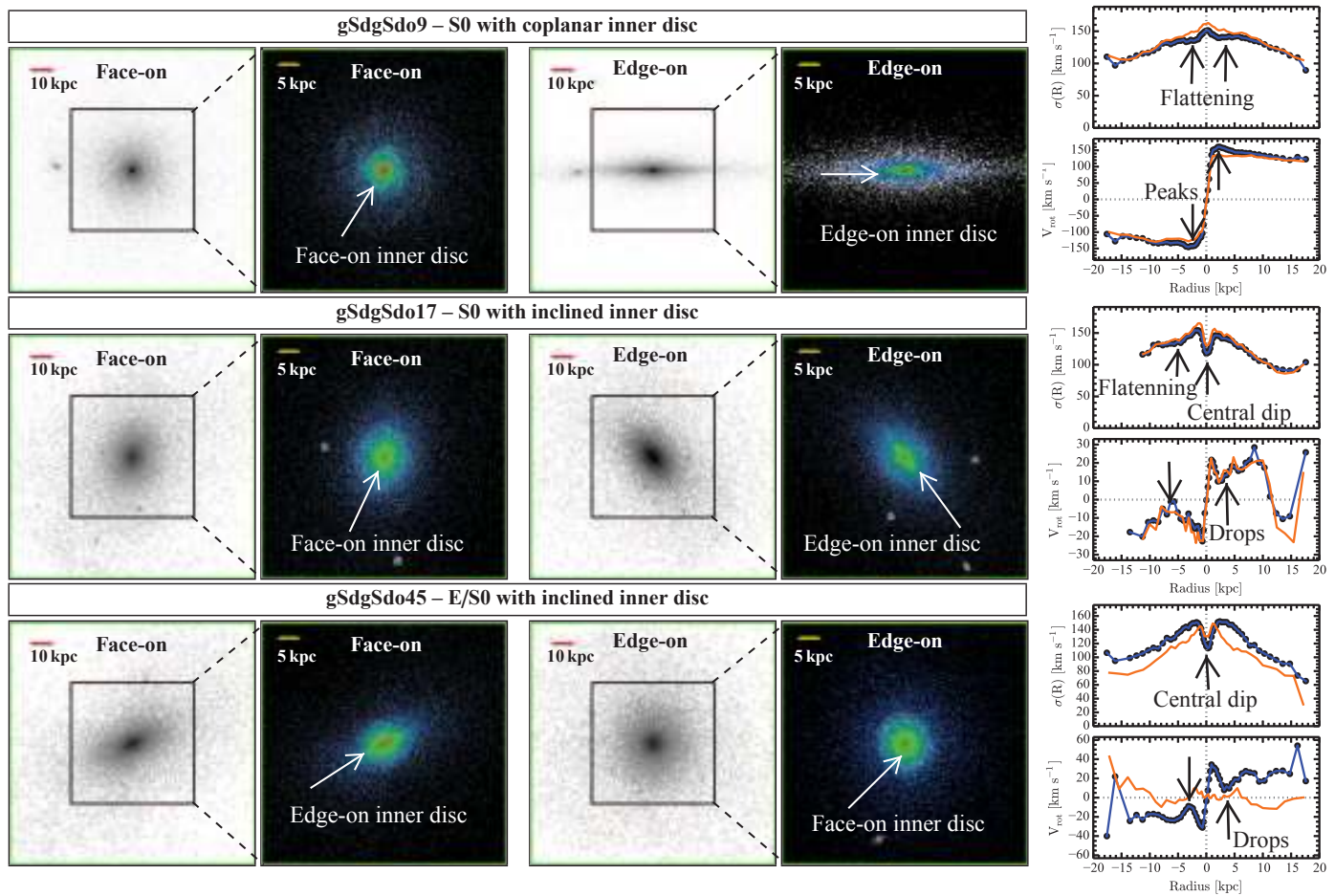

Fig. 12: *V*-band face-on and edge-on images of major-merger S0-like remnants hosting visible young embedded inner discs at their centres. *First and third columns*: *V*-band face-on and edge-on images of each remnant for a FoV of 100×100 kpc², using a logarithmic grayscale that emphasises the flat structure of the inner discs at the centre of the galaxies. *Second and fourth columns*: zooms into the central 50×50 kpc² of the *V*-band images, using a logarithmic staircase colour scale in which the green level follows the inner disc structure. No smoothing has been applied to any frames. *Fifth column*: rotation velocity ($V_{rot}$) and velocity dispersion ($\sigma$) radial profiles of each remnant, both weighted by mass (*orange solid line*) and by *V*-band luminosity (*blue circles*). The inner disc of model gSdgSdo9 is co-planar with the main galaxy disc, whereas those in models gSdgSdo17 and gSdgSdo45 are highly inclined with respect to the main disc (they are more noticeable when seen edge on). The inner disc in gSdgSdo9 is co-rotating with the main disc, because $V_{rot}$ exhibits strong peaks within the innermost ~5 kpc and $\sigma$ becomes shallower in that region (see the luminosity-weighted profile). Those in gSdgSdo17 and gSdgSdo45 are counter-rotating, as shown by the steep decrements in $V_{rot}$ between $R \sim 2$ and $R \sim 5$–10 kpc, where the inner discs are. The $\sigma$ dips at $R \lesssim 1$ kpc in these two last cases are unrelated to these counter-rotating inner discs. They point to even smaller nuclear discs, which co-rotate with the main galaxy disc (notice the strong co-rotation observed at $R \lesssim 1$ kpc in both cases). Any sub-structure with such small sizes are probably affected by the softening of the simulations. For more details on the images, see the caption of Fig. 5.

Concerning the pericentre distance (bottom left panel in Fig. 18), we have obtained lower fractions of S0-like objects as this parameter increases. This trend is even more noticeable in elliptical remnants: the longer the orbital pericentre distance, the lower the number of elliptical remnants formed. Therefore, this behaviour is common to remnants of any types. Encounters in orbits with larger pericentre distances take longer times to merge, so it is reasonable that many of them have not merged at the end of the simulation yet, a fact that can explain the observed trend. However, this trend is also biased by selections effects: first, the number of elliptical remnants in our sample does not account for those with no disc components detectable in their surface density maps, and many of them precisely derive from encounters with large pericentre distances; and secondly, there are lesser experiments in the database with large pericentre distances, so they necessarily represent a lower percentage of the total.

In the bottom right panel of Fig. 18, the percentages of objects are plotted as a function of the spin-orbit coupling. The fraction of merged objects obtained for both kinds of orbits is very similar (48.7% progrades vs. 51.3% retrogrades), so any trends found within the morphological types are unbiased by differences in the number of prograde and retrograde experiments (as occurred with *i*). In fact, we find that the percentages attending to their morphological types differ significantly. The number of S0 remnants is nearly twice more abundant in prograde orbits than in retrograde ones, whereas the opposite is found for elliptical remnants. Accounting for both major and minor mergers and only considering the S0-like remnants in the sample, ∼33% of them derive from retrograde encounters, the majority of them therefore results from prograde ones (∼ 67%). The higher likelihood of resulting in a S0-like remnant found in prograde encounters than in retrograde ones is also coherent with the results of previous studies.

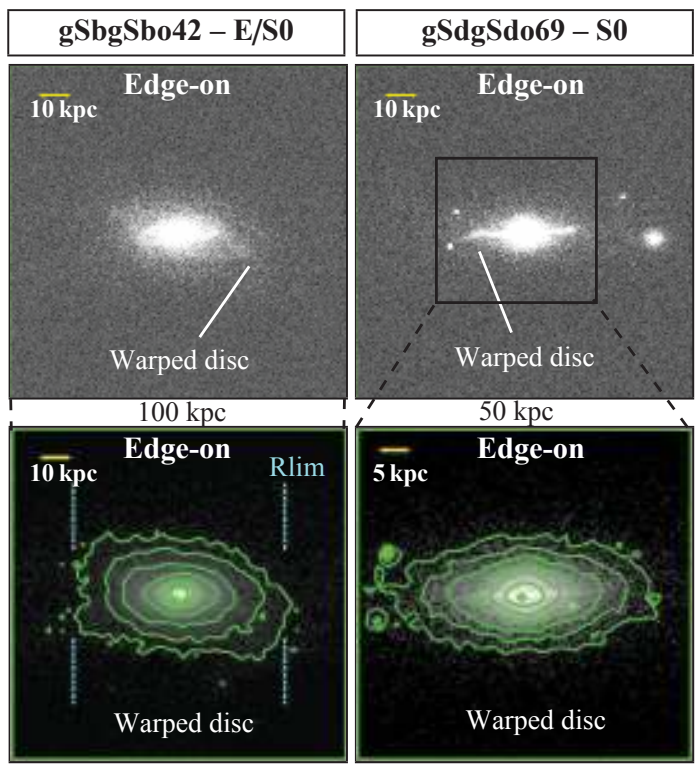

Fig. 13: Edge-on images in the V band of two of the only three S0-like major-merger remnants found to have significant warped discs. The other one is gSbgSbo72, shown in Fig. 7. *Top panels*: original edge-on V-band images of the remnants, for a FoV of $100 \times 100\,\mathrm{kpc}^2$. *Bottom panels*: same images as in the top panels, but applying a Gaussian smoothing to the images with $\sigma = 8$ and 4 pixels respectively (equivalent to 2" and 1"), and zooming into the central $50 \times 50\,\mathrm{kpc}^2$ in model gSdgSdo69. Some isophotes have been over-plotted in the smoothed images to re-mark the bending of the disc in both cases (*green*). In all cases $\mu_{\mathrm{lim}}(V) = 26\,\mathrm{mag\,arcsec}^{-2}$ ($S/N = 5$). For more information, see the caption of Fig. 5.

Although it is true that S0 and E/S0 remants preferentially result from mergers with high initial gas fractions and co-planar prograde orbits (as reported by previous studies), these `GalMer` simulations show that they do not exclusively result from this sort of encounters: S0-like remnants can be formed from a wide variety of initial conditions, including highly-inclined orbits, retrograde encounters, and even progenitors with very low gas contents. We discuss this result in Sects. 9.1 and 9.5.

## 8. Limitations of the models

The limitations of the models have been extensively commented in previous papers of this series (Appendix A in B14, Sect. 5 in Q15a, and Appendices A and B in T17). We describe here only those that may have potential impacts on the morphology of the final remnants, and thus on the present results.

### 1. Initial orbital conditions and progenitor models

The `GalMer` database samples a wide range of initial conditions, but its models are obviously limited by the available computational power and time. This means that they partially cover the enormous diversity of binary encounters that may have occurred through the lifetime of the Universe. In particular, the initial orbits of the encounters are quantified into twelve cases and the numerical models available for the progenitors are limited to five morphologies and two sizes (giants and dwarfs). In fact, these conditions are not intended to reproduce the most probable orbits and merging types observed in cosmological simulations, so many of them must have even been quite rare in the Universe.

Cosmological models have the advantage of analysing the galaxy evolution in a representative volume of the Universe, so they naturally survey the most probable orbital configurations and merging types occurred at each epoch. On the other hand, their spatial and numerical resolutions are usually worse than those of intermediate-resolution binary merger models (as the `GalMer` ones), limiting the study of the formation of ICs with them.

Moreover, the structures of the progenitor models in `GalMer` are quite basic and mathematical, and thus, necessarily unrealistic at some point. Real bulges seem to have any shapes except for spherical (Costantin et al. 2018) and dark matter haloes tend to be triaxial in cosmological simulations (Despali et al. 2014), contrary to the spheroidal shapes assumed for them in `GalMer`. However, numerical studies indicate that the effect of the baryonic matter and dissipation is to make dark matter haloes more spheroidal (Springel et al. 2004; Kazantzidis et al. 2010; Zhu et al. 2017; Sameie et al. 2018), although there are discrepant opinions concerning this (e.g. Athanassoula et al. 2013). Some studies even claim that the effects at the centre of the Milky Way originated by a non-spheroidal dark matter halo must be less relevant than those coming from the uncertainties in the dark matter density profile (Bernal et al. 2014).

This means that we must interpret with care any results which can be directly affected by the shape assumed for the initial mass distribution of the progenitors, such as the lack of large bars in the S0-like remnants from major mergers or the low number of significant warps found among the major-merger sample (see comments in Sect. 9.4). However, this limitation does not affect the main conclusion of this study: the feasibility of building a realistic S0 in terms of morphology from a major merger (see comments in Sects. 9.1, 9.3 and 9.5).

### 2. Spatial and numerical resolutions

`GalMer` models have a spatial resolution limited by the softening length used in the simulations (∼200 or 280 pc). Therefore, they can be used for studying properties in scales from one to several kilo-parsecs, as demonstrated in many papers that have used them (see Di Matteo et al. 2008, 2009a,b; Qu et al. 2011b; Sil'Chenko et al. 2011; Randriamampandry et al. 2016). In smaller regions, they can be affected by the artificial smoothing applied by the code to the gravitational forces. As all features characterised in this paper are larger than ∼1 kpc, they can be considered robust in this sense, in particular, the formation of ovals and lenses from the merger itself (and not through bar dilu-tion) and the lack of large bars in major-merger remnants (they are commented in Sects. 9.3 and 9.4). In fact, the nuclear bars that we have identified visually represent a limiting case of reliable features, because their sizes are ~1–2 kpc (see Fig. 9). The only features identified in the present paper that are really uncertain due to this fact are nuclear compact sources, as already warned in Sect. 6.

Concerning the numerical resolution of the simulations, it is high enough to avoid significant numerical heating, as already proven by studies that analyse the vertical disc thickening produced by these mergers (see Qu et al. 2011a). The number of particles used in them seems enough to reproduce some large-scale phenomena, such as tidal relics, warps, and flares in the remnant discs (as shown here), although we have the limitation of the grainy structure of the outer discs (Sect. 4.1.3). Nevertheless, these simulations are capable of realistically reproducing quite complex large structures, such as the formation of anti-truncations in the remnant discs that reproduce the tight scaling relations found in their real analogues (B14). In any case, the experience with similar simulations in other studies (at least concerning numerical resolution) indicates that the number of particles used in these simulations are adequate for the spatial ranges of the features analysed here (see, e.g. Sellwood & Merritt 1994; Combes et al. 1999; Sellwood & Binney 2002; Revaz & Pfenniger 2004; Shen & Sellwood 2006; Kaufmann et al. 2007; Kim et al. 2014; Peschken & Łokas 2018).

### 3. Mass ratios and progenitor masses

Considering the original masses of the progenitors and mass ratios of the encounters (see Tables 1 and 2), these simulations generate quite massive galaxies ($M_* \geq 3 \cdot 10^{11} M_\odot$). Consequently, their morphological properties should be compared with those of S0 and E/S0 galaxies with similar masses, as done here with the NIRS0S sample (L10; L11).

Obviously, the mass of gas in the progenitors would be much lower if they had been intermediate-mass galaxies or dwarves instead of giants (for the same morphological types). The gaseous mass determines the SFH induced during the interaction. If the gas consumption is proportionally lower, the dissipation of energy and angular momentum will also decrease. The gas is known to play a key role transferring angular momentum between the inner and outer regions of the remnants. In fact, dissipative effects contribute to smooth the global morphology and to form disc components (Jesseit et al. 2007; Hopkins et al. 2009b; Lotz et al. 2008, 2010a). Therefore, the final morphology of a remnant resulting from a gSa+gSa experiment would not necessarily be similar to that resulting from an iSa+iSa or dSa+dSa simulation with the same initial conditions and mass ratios, even though these progenitors are small replicas of the gSa, just because dissipative processes are not scalable. Moreover, in these cases, the times of full merger would also differ, even using the same initial orbital conditions. Therefore, there is still a wide range of mass to explore in this scenario, particularly at low masses. This would be especially interesting, because many studies indicate that the number of mergers undergone by a galaxy during its lifetime increases with the mass ratio of the encounter (although this strongly depends on the mass range too, see Khochfar & Burkert 2001, 2003; Maller et al. 2006; Bournaud et al. 2007b; Wilman et al. 2013; Tapia et al. 2014).

There are other studies that analyse wider ranges of mass ratios than those used here (e.g. Naab & Burkert 2003; Bournaud et al. 2004, 2005b; Eliche-Moral et al. 2006;

Di Matteo et al. 2007; Jesseit et al. 2007; Eliche-Moral et al. 2011). Nevertheless, the fact that we find a significant number of realistic S0 remnants resulting even from extreme cases of major mergers (between giants and in 1:1 encounters) demonstrates that major mergers cannot be excluded from the formation scenarios of S0s, at least attending to their morphological properties. Although several authors have studied the effect of different mass ratios specifically in the morphology of merger remnants (Bournaud et al. 2005b; Jesseit et al. 2007; Lotz et al. 2008, 2010b) and in their final gas and SF amounts (Di Matteo et al. 2007; di Matteo et al. 2008; Cox et al. 2008), this is the first study (to our knowledge) that analyses in detail the formation of morphological features in the remnants that are characteristic of real S0 galaxies.

### 4. Dissipative effects

As commented in the previous item, dissipative effects can strongly affect the morphology of the final remnant. In simulations, their effects are introduced through some numerical recipes that are an approximation to reality. Calibrating them using observational data makes them more reliable, as done in `GalMer` models (see, e.g. Di Matteo et al. 2007; di Matteo et al. 2008).

The gas fractions of the `GalMer` progenitors reproduce the typical values of each morphological type in the local Universe. But gas fractions were much higher in galaxies at the epoch at which the merging mechanism was really relevant (at $z \gtrsim 0.8$, see e.g. Eliche-Moral et al. 2010; Davidzon et al. 2013; Prieto et al. 2013; Choi et al. 2014). As commented before, higher gas amounts trigger the formation of disc components in the remnants, increasing the likelihood of producing an S0 instead of an elliptical. This even strengthens the relevance of major mergers as a mechanism of S0 formation in the past suggested by the present results. We discuss this in more detail in Sect. 7.1.

The gas mass in the progenitors does not only affect the global morphology of the remnant (in the sense of having or not a disc), but it is also crucial for the formation of inner structures. In particular, the gas mass affects to the level of SF in the disc, induces the formation of inner sub-components at the centre through merger-induced starbursts, and makes the main disc more unstable to bars, spiral patterns, or warps. This is also commented in Sects. 9.3 and 7.1.

Finally, higher gas amounts tend to produce longer-lasting tidal relics because: 1) gas particles are extremely sensitive to gravitational torques, and contribute to the formation of long tidal tails (Bournaud et al. 2008a; Knierman et al. 2012), and 2) gas clumps in these tidal tails can collapse and form stars, increasing the luminosity of the tidal relic (Bournaud et al. 2008b). Both facts contribute to make tidal structures more detectable under similar observing conditions.

### 5. Mass-into-light conversion and total simulation time

Section 4.1 describes the procedure we have used to assign a $M/L$ value in a given band to each stellar particle in the remnants. We obviously had to adopt several assumptions to reduce the complexity of the physical system. But different diagnostics showed that the global visual morphology of the final remnant was extremely robust to the assumptions adopted in this procedure, we mean its visual morphological type did not change significantly.

However, we also found that some ICs could be masqueraded or emphasised in the images depending on the age assumed for the old stellar material, especially in the blue bands.

As commented in Sect. 5, many of these ICs result from central starbursts induced during the interaction, so they are made of stellar populations with $\lesssim$1–2 Gyr. We have assumed that all collisionless stellar particles are $\sim$10 Gyr old, independently of their location in the original progenitor and its morphological type, so these young ICs usually out-shine the underlying old stellar light distribution. But this is an over-simplification of reality, because even in S0 galaxies, the average age of their bulges is usually different from that of their discs, and in both cases, these ages can expand a wide range of values (see Sil'chenko 2006; Silchenko & Afanasiev 2008; Sil'chenko et al. 2012; Katkov et al. 2015). Assuming a slightly younger age for old stellar particles (for example, $\sim$7 Gyr) increases the brightness of the whole remnant structure, hiding many embedded young ICs in the global light distribution.

Moreover, these young inner structures would fade and redden enough to become undetectable if the remnants were evolved for $\sim$1–2 Gyr more (see Sect. 5). Therefore, the total simulation time is also critical for their visual detection. Time would also contribute to relax even more the global structure of the remnant and make tidal features much fainter and difficult to detect.

### 6. Dust effects

We have not considered the effects of dust extinction in the photometric images, because this would have required the adoption of too many uncertain assumptions on the distribution and nature of the dust in the remnants, increasing the complexity of the problem unnecessarily. As commented in Sect. 4.1.1, the dust is associated with the hybrid particles in the simulations, which tend to concentrate at the remnant centre and have transformed most of its original gas mass into stars by the end of the simulation. The remaining gas mass is extremely low, see, as it corresponds to a realistic S0 or E/S0 galaxy (Eliche-Moral et al., in prep.). Therefore, dust extinction and reddening could only be significant at the core, but they must be extremely low in most of the galaxy body. This could affect the visual detection of some ICs, but not the global morphology of the remnant.

## 9. Discussion

### 9.1. Remnants with realistic S0 morphologies

`GalMer` models specifically demonstrate that major mergers can produce remnants with realistic and relaxed S0 morphologies in less than $\sim$3 Gyr. Despite the fact that numerous recent studies support this scenario, in other words that major mergers must have generated remnants with discs in many cases and, in particular, S0s (e.g. Costantin et al. 2018; Hammer et al. 2018; Li et al. 2018; Méndez-Abreu et al. 2018, among many others), the major-merger mechanism has been systematically discarded for S0 galaxies, when not directly ignored, in studies specifically dealing with the properties of S0s. They are traditionally pictured as destructive events that can hardly form relaxed S0s (see Sect. 1). However, the results obtained here and in these other studies urge to change this view.

The key difference between other studies and the present one is the selection of the remnants on the basis of their visual morphologies in mock photometric images, instead of using other representations which do not account for the different $M/L$ values of the stellar populations or the limitations inherent to observational data (see Sect. 3). This approach has allowed us to reject a big set of models from the `GalMer` database that, despite resulting in remnants with a disc component in projected density maps, looked like elliptical galaxies in realistic photometric images. This analogy of models and real data is fundamental to perform fair comparisons between them and elucidate the feasibility of the merging mechanism to generate S0 galaxies.

Here we have demonstrated that major mergers, even in extreme cases of 1:1 events, can result in remnants with morphological properties consistent with real S0 galaxies. And this is not exclusively associated with co-planar encounters or huge gas amounts in the progenitors (Sects. 7.1 and 7.4). Therefore, we can expect that mergers with higher mass ratios will result in S0 galaxies more easily, as the progenitor discs will experience less destructive effects. Moreover, these encounters will take longer to merge, so the orbit of the encounter will experience a higher circularisation. This will probably make the progenitors finally merge at lower orbital inclinations than initially, favouring the survival of discs, as already observed in minor-merger simulations (Eliche-Moral et al. 2006, 2011).

### 9.2. Lack of morphological traces of past merger activity

The analysis of the morphologies of the remnants using mock images has also demonstrated that, contrary to the popular expectation, the galaxies deriving from major mergers do not necessarily exhibit significant traces of their violent formation after relaxing times of just $\sim$1–2 Gyr (Sect. 7.3). Most S0 and E/S0 remnants resulting from major mergers in our sample ($\sim$76%) are completely relaxed and do not exhibit any morphological features pointing to their past merger origin (not even small ones) $\sim$1–2 Gyr after the full merger, for typical observing conditions and a relatively nearby distance.

On the other hand, we have detected significant traces of past merger activity in the remnants of the gS0+dwarf experiments ($\sim$ 66%). Although this is also a direct consequence of the shorter relaxing time periods imposed to minor mergers than to major ones (Sect. 3.3), it is also true that the distortions produced by minor mergers are much more persistent than in major ones, despite being more gentle events. These facts should start to be seriously accounted for, since the role of major mergers is systematically discarded in galaxies without catastrophic tidal relics, whereas minor mergers are always contemplated when faint tidal traces are detected (see references in Sect. 1), contrary to the results obtained here.

Duc et al. (2015) performed a census of the fine structures (tidal tails, streams, distortions, and shells) around 92 early-type galaxies from the ATLAS[3D] sample, using very deep $g$-band images (deeper than our mock $R$-band images by $\sim$1 mag). They preliminarily found that $\sim$35% of their sample exhibited a relaxed appearance with no merger relics in deep images, $\sim$22% were interacting, $\sim$16% showed minor-merger relics, and $\sim$12% had clear signs of having undergone a major merger. The criteria adopted by Duc et al. clearly identified as major mergers only those remnants at early post-merger stages of the encounter, soon after the full merger[4] (see their Figs. 19 and 24). Their criteria to identify galaxies that have experienced a major event is very reliable exclusively for recent events. But then any results concerning the major-merger origin of the early-type galaxies in this study must be exclusively limited to the events that may have occurred during the last $\sim$3–4 Gyr at most because, according to the results found here, the relics of major mergers occurred at earlier epochs have probably weakened or diluted with time as to be confused with minor-merger relics according to their criteria or have become undetectable even in data as deep as theirs.

---

[4] In the `GalMer` database, there are remnants with morphologies as distorted as those identified as major-merger remnants by Duc et al., but we intentionally discarded them to find out whether major mergers could really reproduce the regular and relaxed morphologies observed in present-day S0s (see Sect. 3.2).

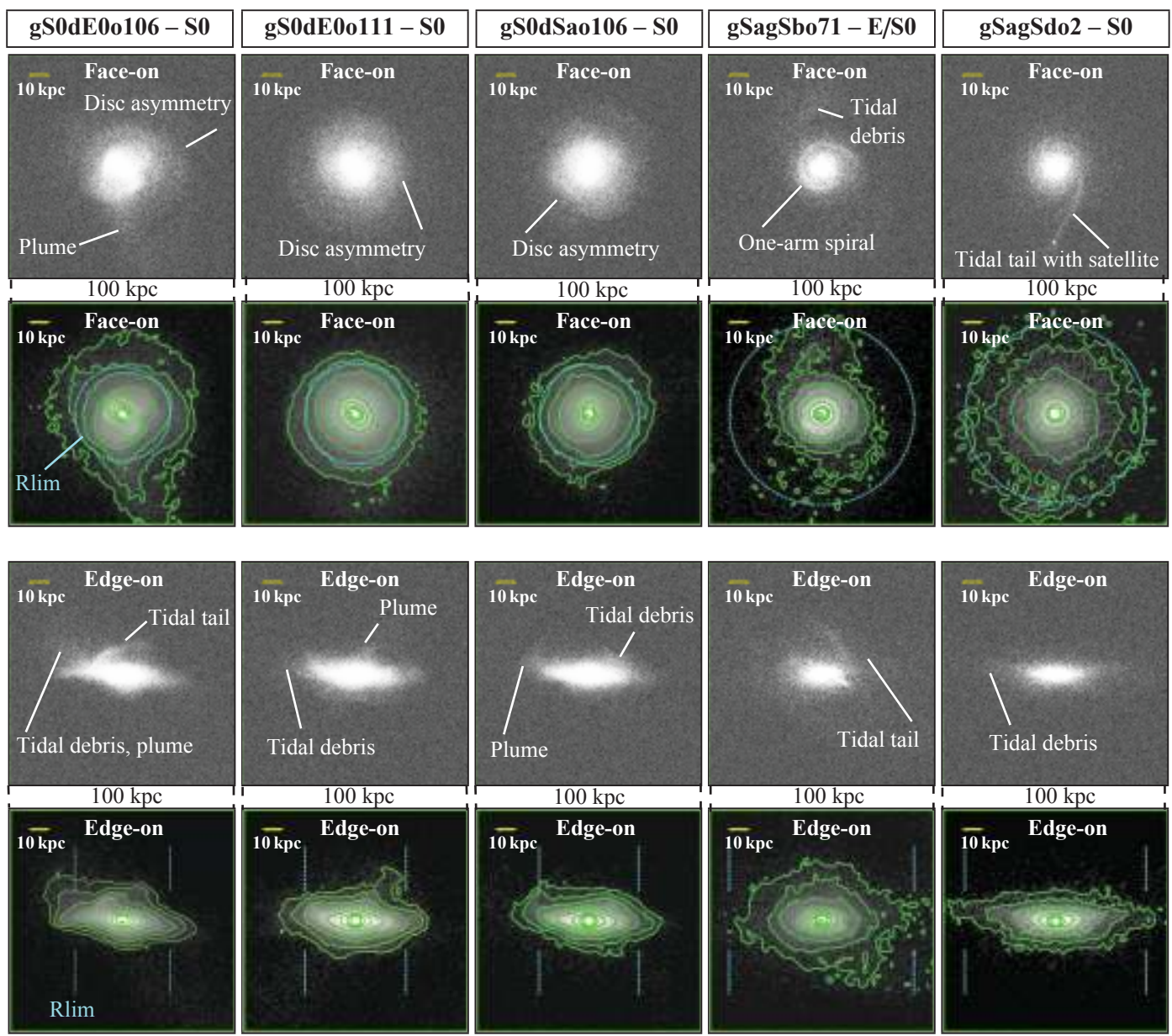

Fig. 14: Face-on and edge-on *V*-band images of S0-like remnants exhibiting some traces of past merger activity. The FoV is $100 \times 100$ kpc$^2$ and $\mu_{\text{lim}}(V) = 26$ mag arcsec$^{-2}$ ($S/N = 5$) in all cases. *First and third rows of panels*: *V*-band images of the remnants in face-on and edge-on views with no smoothing, using a logarithmic grayscale that saturates in the centre to emphasise the sky noise and the fainter structures at the outskirts. Some tidal features that can be detected in each remnant are indicated in the panels. *Second and fourth rows*: same images as in the rows above, but using a Gaussian smoothing with $\sigma = 10$ pixels (2.5"). We have represented some isophotes (*green*) to emphasise the shape of the tidal features indicated in the panels above. The location of $R_{\text{lim}}$ is also marked to indicate the region at which the emission of the galaxy disc falls below $S/N = 5$ in the image. Tidal tails in major-merger remnants tend to be small (e.g. models gSagSbo71 and gSagSdo18). One exception is the remnant of model gSagSdo2, which also contains a tidal satellite at the end of the tail. Tidal debris and plumes are observed in many models (e.g. gS0dSao106 or gSagSdo70). We have found one collisional ring (gSbgSbo70). In general, merger relics are more frequently detected in S0 remnants of minor mergers than in those of major ones for similar time evolution periods (66% vs 24%). See the caption of Fig. 5 for more details. [**Note.**– *The figure continues.*]

There is a trend to believe that the strong tidal relics that appear soon after the full merger in a major encounter will remain for a long time period in the remnant. But many studies indicate that is not necessarily true (see, e.g. Lotz et al. 2008, 2010a,b).

Considering that a galaxy can only have resulted from a major event when strong interacting features are found in it (such as shells, long tidal tails, or train-wreck morphologies) is clearly biased towards recent events, because these signs are only detectable during a relatively short period of time soon after the full merger. More advanced stages of major-merger remnants may not exhibit any strong tidal relics in real (deep) data just ~1–

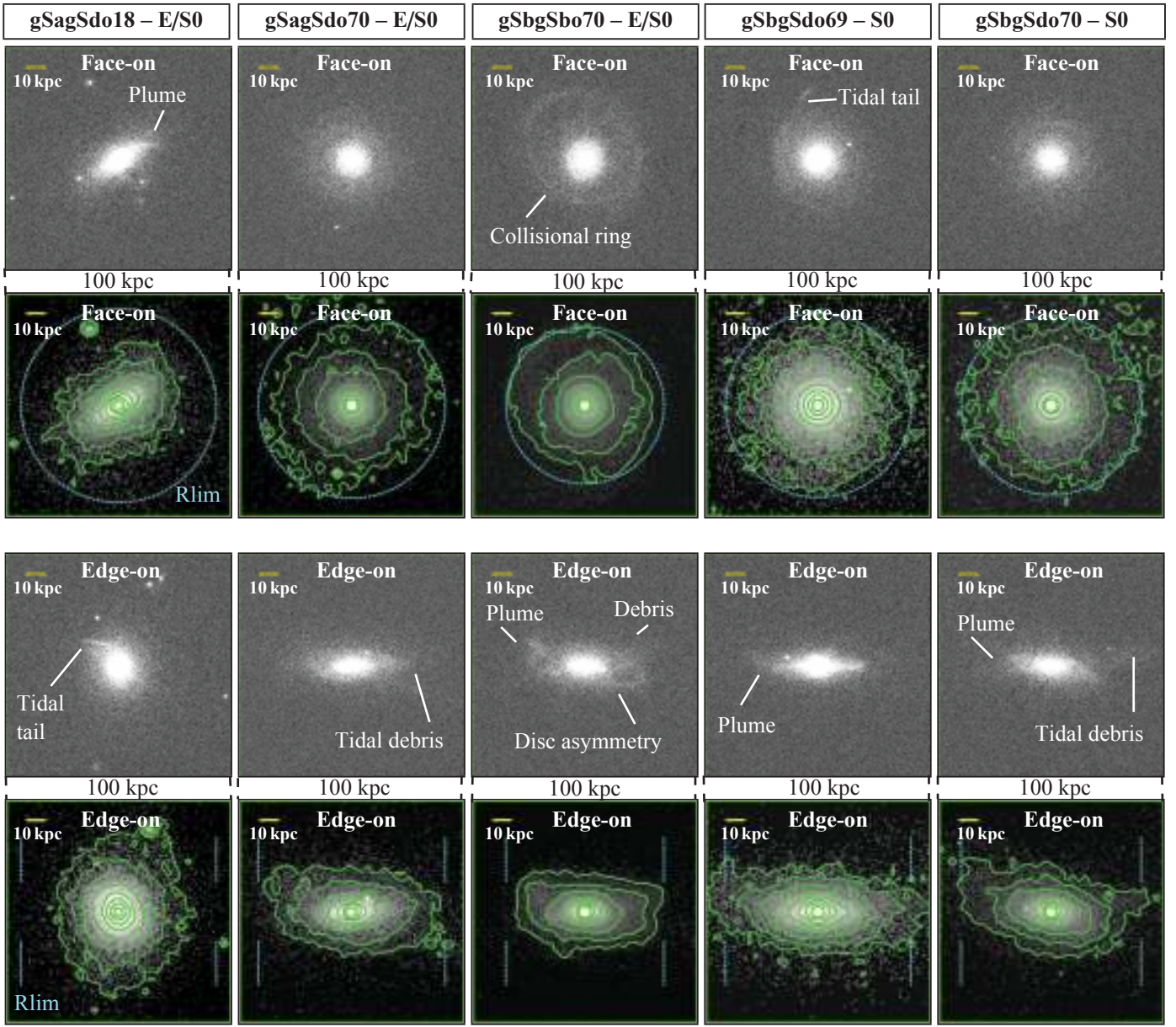

Fig. 14: [Cont.]

2 Gyr after the full merger, as observed here. This means that the interpretation of the tidal features in this kind of observational studies as probes of different past events (major or minor, wet or dry mergers) must be limited only to the mergers occurred during the last ~3–4 Gyr of the cosmic history, but not beyond (see also Atkinson et al. 2013; Hood et al. 2018; Morales et al. 2018).

Therefore, the traditional view of major mergers as events of long-lasting catastrophic effects is not realistic at all. Major mergers can be considered violent events in the sense that they can rebuild a new galaxy, quite different from the original progenitors, but the resulting galaxy can suffer such strong relaxation that it can exhibit an ordered appearance and no traces of its past merger-origin just a couple of giga-years after the event in deep data. So, contrary to popular expectations, the lack of merger-related features and the relaxed appearance of many present-day S0s cannot be interpreted as direct evidences of the poor role played by major mergers in the buildup of this galaxy population.

### 9.3. Ovals, lenses, inner discs, and other ICs

As commented before, the majority of early-type galaxies host ICs, being much more frequent in S0s than in spirals. If major mergers are plausible mechanisms to form S0 galaxies, then their remnants must tend to exhibit ICs. We have checked with the models that this really happens: most of the S0-like remnants in our sample exhibit ICs in their photometric images, such as ovals, lenses, inner rings, pseudo-rings, inner discs, inner spirals, nuclear bars, and compact sources (Sect. 7.2).

We consider significant that, in particular, all S0 and E/S0 remnants host a lens or oval, in agreement with observations (~97% of real S0s host one, see references in Sect. 7.2). In fact, although the remnants from the major mergers do not develop any relevant bars, they all host a lens or oval, consistently with

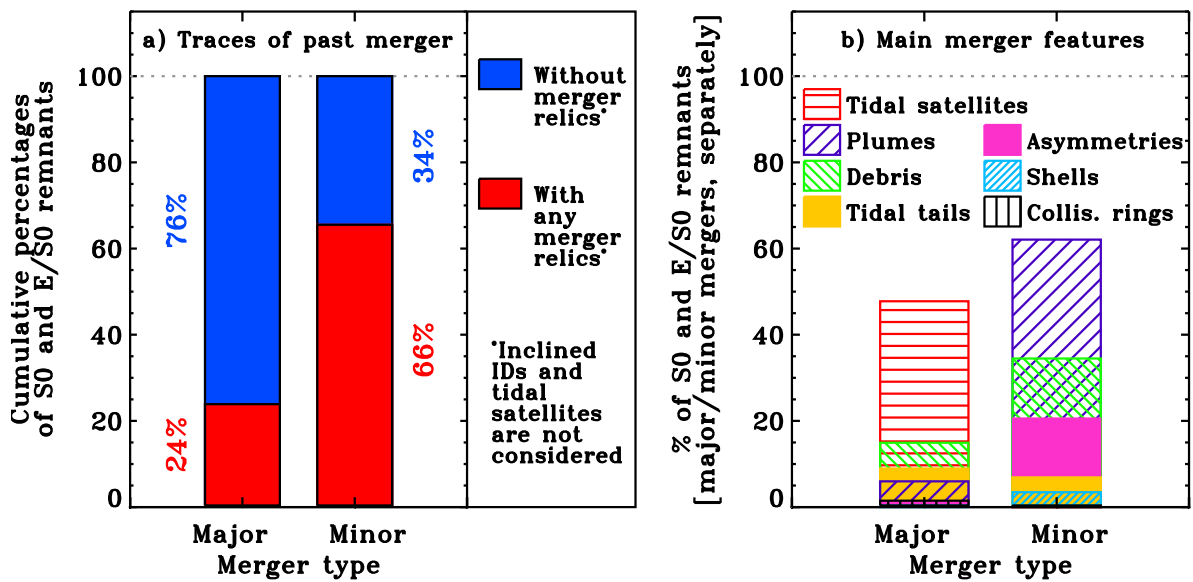

Fig. 15: Percentages of S0-like remnants resulting from major and minor mergers exhibiting merger relics. *Panel a*: cumulative percentages of remnants with and without any merger relics for the major- and minor-merger samples separately. *Panel b*: percentages of some tidal features detected in the S0-like remnants for the major- and minor-merger samples. See the corresponding legend in each panel. More information in Table A.5.

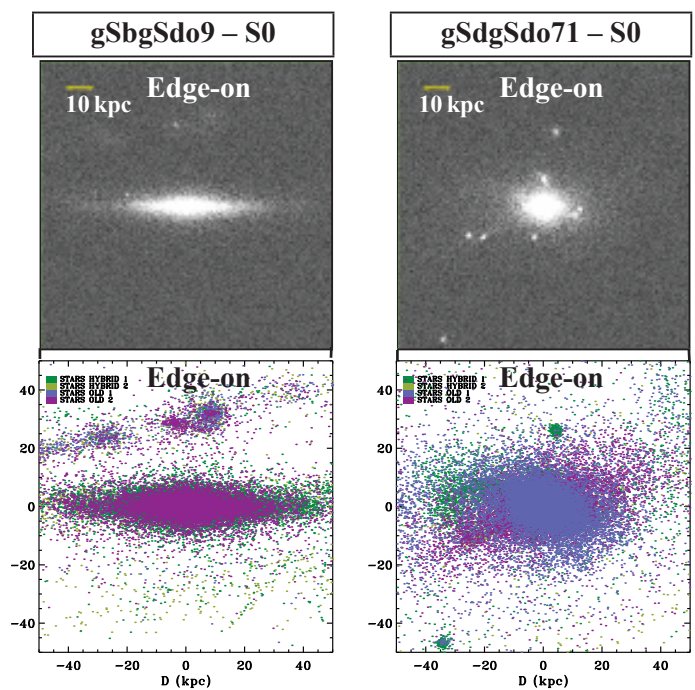

Fig. 16: Tidal satellites formed in a couple of S0 remnants in our sample: gSbgSdo9 (*left panels*) and gSdgSdo71 (*right panels*). We represent their edge-on *V*-band images (*top*) in comparison to their maps of projected particle positions (*bottom*) for the same points and FoVs to show how observational effects can masquerade the tidal nature of these satellites. No smoothing has been applied to the images. For more information, see the caption of Fig. 3.

the observational fact that a large majority of unbarred S0s also host ovals or lenses (L09; L11; L13). Most observational studies attribute these features to bar dilution and evolution (Sil'chenko

2002; Laurikainen & Salo 2016, 2017; Buta et al. 2010, L09; L13). However, these simulations demonstrate that lenses and ovals are not exclusive of internal secular evolution (in particular, of bar dilution), but they can result from a merger as well. This should be taken into account when interpreting observational data.

Similarly, other ICs such as inner discs, inner spirals, inner pseudo-rings, rings, and nuclear bars are usually attributed to gas stripping, secular evolution, gas infall, or minor merging (Emsellem et al. 2004; Kormendy & Kennicutt 2004; Sil'chenko et al. 2011; Laurikainen & Salo 2016; Sil'chenko et al. 2010; Gadotti et al. 2015, L09; L10; L13; KB12), but we have seen here that they can derive from major mergers too, without requiring the formation of any bars (Sect. 7.2). Even though the final remnants are devoid of gas, the dynamical processes and SF triggered by the interactions themselves are capable of producing this sort of sub-structures at the remnant centres, so frequently observed in real S0s. Therefore, the fact that S0s usually contain ICs does not necessarily point to an evolution from spirals through gas stripping or secular processes: major mergers can also produce remnant S0s from spirals with this sort of central sub-structures, even in the case of gas-poor progenitors. Moreover, it cannot be discarded either that ICs can be formed in a S0 after its formation, for example, through later gas infall. So the presence of ICs in a galaxy is not necessarily linked to the process responsible of its main buildup.

### 9.4. Lack of bars in major-merger S0-like remnants

It is striking that none of the S0 or E/S0 remnants coming from a major merger presents a relevant bar in the sample (Sect. 7.1). We have found nuclear bars in many cases (as also observed in real S0s, see Erwin & Sparke 2002; Laine et al. 2002, L09), but none extends beyond ∼1–2 kpc in remnants that have external radii ∼10–25 kpc. The lack of significant spiral patterns, SF, and even large bars in the discs of these major-merger remnants must be interpreted with caution, because all these effects could be due to limitations of the simulations, such as the limited set of initial conditions (in particular, the gas fractions).

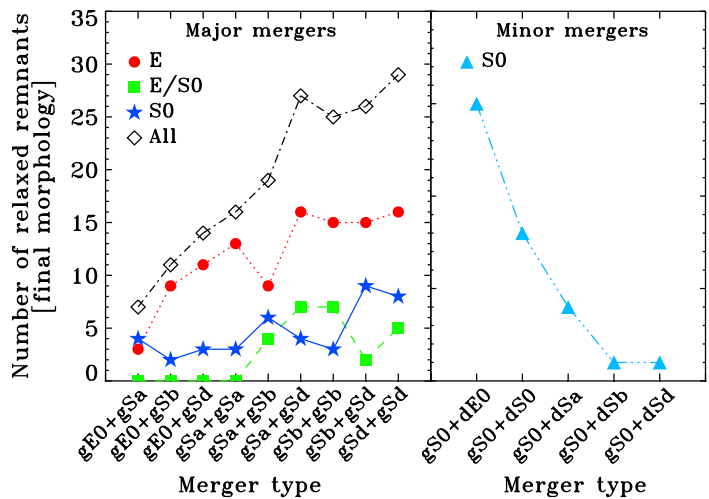

Fig. 17: Number of remnants as a function of the morphological types of their progenitors for each kind of morphological type of the resulting remnant (E, E/S0, S0), for major and minor mergers separately. See the legend in the figure.

The gas fractions in the progenitors of `GalMer` simulations reproduce typical values of local morphological types (Sect. 2), but gas fractions in disc galaxies were much higher in the past (up to ∼50% at $z \sim 1$, see Papovich et al. 2005; Genzel et al. 2008; Tacconi et al. 2008; Förster Schreiber et al. 2009; Law et al. 2009). Therefore, many major mergers that occurred at intermediate-to-high redshifts may have resulted in S0s (and even spirals), instead of ellipticals. Many observational cases at low and intermediate redshifts supporting this idea have been found by Hammer and collaborators (Hammer et al. 2005, 2009b,a, 2010, 2012, 2013; Peirani et al. 2009; Puech et al. 2009, 2012, 2014; Yang et al. 2009; Wang et al. 2015). But also recent theoretical studies point to this. Athanassoula et al. (2016) and Sauvaget et al. (2018) have published major-merger simulations using higher gas fractions in the progenitors to reproduce the conditions of discs at $z \sim 1.5$, and they all resulted in barred spiral remnants.

Although simulations show that tidal encounters tend to trigger non-axisymmetric distortions (bars) in discs (Mihos et al. 1995; Mihos & Hernquist 1996; Bournaud et al. 2005b; Mastropietro et al. 2005; Chan & Junqueira 2014; Moetazedian et al. 2017), we cannot discard the possibility that major mergers really tend to produce unbarred S0 and E/S0 remnants for relatively low initial gas fractions, because mergers are also known to induce bulge growth (Aguerri et al. 2001; Bournaud et al. 2007b; Eliche-Moral et al. 2006, 2012), and massive central spheroids prevent self-gravity and contribute to the stability of galaxy discs against bar distortions at the same time (Pfenniger & Norman 1990; Bournaud & Combes 2002; Athanassoula et al. 2005; Eliche-Moral et al. 2006, 2011; Cox et al. 2008; Athanassoula et al. 2013; Kataria & Das 2018). So, if the resulting remnant bulge is massive enough, then it is probable that the remnant disc is unable to develop any relevant bars. Most S0-like remnants in our sample (∼93%) have bulge-to-total ratios ($B/T$) > 0.3 in the $K$ band (see Q15a), values prominent enough to prevent the formation of spiral and bar distortions in the discs in gas-poor conditions.

We also remark that most discs in our S0-like remnants are relevant components: their disc-to-total luminosity ratios range $0.18 < D/T < 0.56$ in the $K$ band (Q15a). Therefore, they are not the same case as S0(0) galaxies, which are never barred probably because their $D/T$ values are so small that they cannot develop a bar instability (KB12).

This result from simulations could have two observational counterparts: first, the trend of real bulges with high Sérsic indices ($n > 2.5$) to be hosted by unbarred galaxies (∼69%), and secondly, the lack of bulges with triaxial and barred structures found among S0s in comparison to spirals (Costantin et al. 2018). Therefore, although the formation of massive bulges can also be a numerical effect of simulations (Hopkins et al. 2009b), it may be true that the S0s deriving from major mergers between progenitors with gas fractions similar to those of local galaxies (including spirals) tend to be unbarred, as found here.

This result is also supported by the observational fact that the percentage of bars in S0s decreases with the bulge size: from ∼80% in those with small bulges (S0+) to ∼35% in those with big ones (S0−, see L13). Moreover, the percentage of bars in S0s decreases from ∼24% in faint S0s ($M_K > −24.5$) to 6% in bright ones[5] ($M_K < −24.5$, Barway et al. 2011, see also L13). Both trends (with morphology and luminosity/mass) are coherent, considering that bright (massive) S0s tend to host the bigger (more classical) bulges (Kormendy & Kennicutt 2004; Vaghmare et al. 2013, L10). If we consider the existence of more classical bulges in massive S0s as a sign for the higher relevance of mergers in their formation, then it is reasonable to find a lower fraction of bars among bright S0s too, in agreement with these simulations.

The fraction of bars found in local S0s strongly depends on the detection method used to identify them, but conservative methods establish a lower limit of ∼12–15% for it (Nair & Abraham 2010; Barway et al. 2011). Studies including both strong and weak ones report fractions from ∼46% (Barway et al. 2011, L13) up to ∼60% (de Vaucouleurs 1963; Aguerri et al. 2009), showing no significant trends with environment among bright S0s (Marinova et al. 2012). This means that at least ∼40% of S0s do not have any bars (strong or weak). Therefore, the fact that the S0-like remnants coming from major mergers do not develop significant bars cannot be used as an argument against the merger origin of S0s in general, as at least ∼40% of real ones do not host any bars.

Considering that the fraction of bars is much lower in local S0s than in spirals (29% vs. 55%, see Aguerri et al. 2009), it can be inferred that the processes that have driven the evolution of spirals into S0s must also inhibit bars in many cases (see also L09). Mechanisms that simply remove gas from the galaxy (such as gas stripping, strangulation, or starvation) may not be capable of explaining the total disappearance of stellar bars from the spiral progenitors to their S0 descendants in time periods shorter than ∼2 Gyr, because stellar bars (which would not be affected by these processes) seem to be very stable structures in gas-poor scenarios (Debattista et al. 2004, 2006; Berentzen et al. 2007; Villa-Vargas et al. 2010; Athanassoula et al. 2013), although specific simulations on these processes are required. Numerical simulations find that secular evolution itself does not efficiently destroy bars either: bars can weaken as they grow a bar-lens component, but "this weakening is slow and it would have a sizeable effect only at times much longer than a Hubble time" (Athanassoula et al. 2013). On the other hand, the forma-tion of central mass concentrations in galaxies is proven to be an efficient mechanism of bar dilution and its prevention, as commented before. Therefore, it is quite feasible that many massive unbarred S0s may derive from major mergers between galaxies with relatively low initial gas fractions.

---

[5] The S0 and E/S0 remnants in our sample present $M_K < −24.5$ (see Q15a).

Another observational fact supporting a major-merger formation for many massive S0s is that bright S0s in clusters exhibit a higher bar fraction than their counterparts in the field/groups (Barway et al. 2011). Considering that the evolutionary effects of merging have probably been more relevant in groups and the field than in clusters (Sect. 1), this could explain why there are more unbarred S0s in poor environments than in clusters.

Moreover, L13 consider that disc fading and environmental processes alone cannot explain the lower fraction of bars in S0s (~46%) with respect to their most immediate ascendants (~93% in S0/a galaxies and ~64% in spirals). As this is compensated by a higher fraction of lenses and ovals in S0s compared to spirals (97% vs. 82%), these authors conclude that the strong bars present at spirals must have weakened into the ovals and lenses we observe in present-day S0s. The bar weakening or dilution mechanism cannot be discarded and must have occurred in many cases (although see the discussion on the lifetimes of bars provided in Sect. 5 of Athanassoula et al. 2013). In any case, the present simulations demonstrate that both the lower fraction of strong bars and the higher fraction of lenses and ovals in S0s with respect to spirals can also be explained through a major-merger origin (see Sects. 7.1 and 7.2).

Concerning minor mergers, several events of this kind would probably be required to turn a gas-rich galaxy into an unbarred S0, because we have seen that, if the progenitor initially had a bar, the satellite accretion destroys it with some difficulty (Sect. 7.1). On the contrary, some studies indicate that minor mergers tend to trigger bars in galaxy discs, even in the absence of gas in the progenitors (Eliche-Moral et al. 2011, and references above). So minor mergers do not seem to be good candidates to explain the transformation of spirals (mostly barred) into unbarred S0s. They would probably lead to barred S0s instead.

We cannot exclude either that many bars in S0s can have been formed through gas infall after the formation of the galaxy itself through a major merger. Prograde encounters efficiently eject gas to the environment at early stages of the merger, so this gas can fall onto the remnant later again (Di Matteo et al. 2007). If gas infall is invoked to explain the formation and renewal of bars in spirals (Bournaud & Combes 2002; Bournaud et al. 2005a), then it can explain the formation of a new bar in an unbarred S0 galaxy that has resulted from a major merger too. Moreover, many unbarred S0s may obviously derive from unbarred spirals that have lost their gas through mechanisms different from mergers (Rizzo et al. 2018). Nevertheless, these simulations show that major mergers can provide a feasible explanation for the transformation of many barred spirals into their unbarred S0 descendants.

### 9.5. Feasibility of the merger mechanism for S0s

We have found 67 S0-like relaxed remnants out of the 876 major merger experiments initially in the database, thus ~8% of the encounters produce a relaxed S0 or E/S0 galaxy. This percentage may be higher, because many models are not merged and relaxed at the end of the simulations, after a total computed time period of ~3–3.5 Gyr. Although the orbital parameters of GalMer experiments are not intended to reproduce the most frequent orbital configurations found in cosmological simulations, they are diverse enough to infer that the likelihood of a major merger to result in a S0-like galaxy (instead of an elliptical) is quite significant, even for gas fractions typical of the present-day Universe.

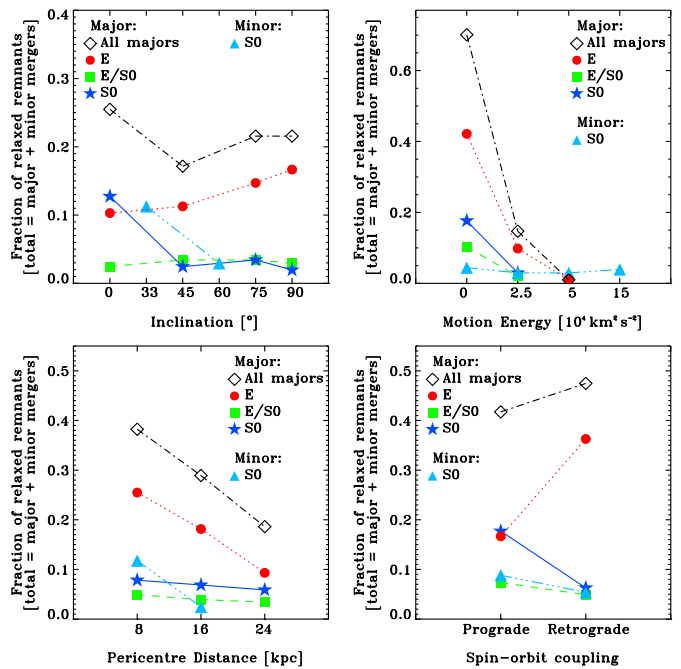

Fig. 18: Fractions of remnants of each morphological type within the sample of 202 relaxed remnants (including both major and minor mergers) as a function of the initial orbital parameters of the simulations: orbital inclination, initial kinematic energy, pericentre distance, and spin-orbit coupling. See the legend in the panels

Accounting for the fact that spirals had higher gas fractions in the past (see references above), this percentage would probably be much higher, because higher gas amounts contribute to disc rebuilding and preserving, as these simulations and previous studies confirm (see Sect. 7.4 and references in Sect. 7.4). Additionally, considering that cosmological simulations also indicate that low-inclined ($i < 40^\circ$) prograde encounters were more frequent in the early Universe (Knebe et al. 2004; Zentner et al. 2005; Gómez-Flechoso et al. 2010; Benjouali et al. 2011) and that this sort of orbits result more easily into S0-like remnants than highly-inclined ones (Sect. 7.4), the likelihood of a major merger to result in a S0 galaxy along the lifetime of the Universe must have been relevant enough to consider major mergers as a feasible formation mechanism for many massive S0s.

The appearance of the S0 population at the expense of the spirals since $z \sim 1$ is an observational fact (Dressler 1980). But we are still far from elucidating the real role played by the different evolutionary mechanisms in the transformation of spirals into S0s, and in particular, of mergers. Many studies indicate that the majority of the massive E-S0 population must have undergone at least one major merger since $z \sim 1$–1.5 (Eliche-Moral et al. 2010; Prieto et al. 2013; Prieto & Eliche-Moral 2015; Tapia et al. 2014; Rodríguez-Gómez et al. 2016). This suggests that a significant fraction of discs must survive or be rebuilt after major mergers to give rise to some S0 galaxies.

Therefore, contrary to the widespread idea that major mergers cannot produce normal S0 galaxies in a short time period, these GalMer simulations prove that major mergers between spirals can result in S0 and E/S0 remnants with morphological properties consistent with local S0s in ≤3 Gyr, for diverse orbital configurations and even in gas-poor conditions. This contributes to reconcile the observed properties of S0s (mainly of massive ones) with hierarchical scenarios of galaxy formation, although we must stress that the merger origin of many S0s does not exclude the relevance of other mechanisms in the buildup of this galaxy population either, mainly those in clusters and with low masses (Calvi et al. 2012).

## 10. Conclusions

The role of major mergers in the formation of massive S0 galaxies is still controversial. The relaxed and regular structures shown by most present-day S0s and the ICs that they frequently host (ovals, lenses, inner discs) are considered incompatible with a major-merger origin, being interpreted as signs of a formation through internal secular evolution, gas infall, or environmental mechanisms instead.

In order to elucidate the feasibility of the merging mechanism to produce remnants with realistic S0 and E/S0 morphologies, we have analysed the morphological properties of a set of dynamically relaxed remnants resulting from the SPH N-body simulations of major and minor mergers available in the `GalMer` database. We have visually identified the remnants with S0-like morphology using mock photometric images that reproduce the typical observing conditions of current local surveys and compared their global appearances, visual morphological components, and merger relics with those of local S0 galaxies. We have found the following results:

1. Contrary to the traditional view of major mergers as catastrophic events, incapable of producing S0 galaxies, major mergers can produce S0 and E/S0 remnants with realistic relaxed morphologies in ≤3 Gyr.

2. Most S0-like remnants show no trace of their past merger origin under typical observing conditions for relaxing periods ≤1–2 Gyr, at distances as nearby as only 30 Mpc.

3. The merger traces are more persistent and easier to detect visually in the S0-like remnants of minor mergers than of major ones for similar relaxing time periods.

4. Although S0-like remnants result more probably from mergers with high initial gas fractions and in co-planar prograde orbits (as already reported in previous studies), S0 and E/S0 remnants are formed in mergers with quite diverse orbital configurations and low initial gas amounts.

5. The majority of the S0-like remnants host ICs such as ovals, lenses, inner discs, inner rings and pseudo-rings, inner spirals, nuclear bars, and compact sources, all frequently found in real S0s too.

6. Edge-on views of these S0-like remnants reveal that their ovals and lenses are associated with stellar halos or embedded (thin or thick) inner discs resulting from the merger, and not with diluted bars (as usually assumed).

7. In fact, no major-merger S0-like remnant in the selected sample develops a significant bar. This result might be quite significant, as major mergers could then explain why ≳40% of real S0s are unbarred, whereas their immediate progenitors are mostly barred (~93% in S0/a galaxies and ~64–69% in spirals), a fact difficult to explain with other scenarios of S0 formation, such as disc fading, environmental mechanisms, or even minor mergers.

`GalMer` simulations demonstrate that major mergers can produce remnants with morphologies consistent with those of real S0s. Therefore, we conclude that the morphological properties of present-day S0s do not exclude major mergers as a feasible formation mechanism at all, specially in massive S0 systems.

*Acknowledgements.* The authors thank the anonymous referee for comments that have significantly improved this manuscript. We are grateful to I. Chilingarian, P. Di Matteo, F. Combes, A.-L. Melchior, and B. Semelin for creating the `GalMer` database and the HORIZON project for supporting it (http://www.projet-horizon.fr/rubrique3.html). MCEM thanks P. Di Matteo and E. Laurikainen for their kind support concerning the `GalMer` database and the properties of galaxy lenses, respectively. We also acknowledge the usage of the HyperLeda database (http://leda.univ-lyon1.fr). This research has made use of the NASA's Astrophysics Data System and the NASA/IPAC Extragalactic Database (NED) which is operated by the Jet Propulsion Laboratory, California Institute of Technology, under contract with the National Aeronautics and Space Administration. This research has also made use of SAOImage DS9, developed by Smithsonian Astrophysical Observatory.

## Appendix A: Tables miscellanea

Table A.1: Orbital parameters of the `GalMer` models that result in dynamically relaxed remnants

| Orbit ID (1) | Spin-orbit [P/R] (2) | $i_2$ [°] (3) | $D_{Per}$ [kpc] (4) | $E_0$ [$10^4\,\mathrm{km^2\,s^{-2}}$] (5) | No. models (6) |
|---|---|---|---|---|---|
| 1 | P | 0 | 8 | 0 | 4 |
| 2 | P | 0 | 8 | 2.5 | 4 |
| 5 | P | 0 | 16 | 0 | 8 |
| 9 | P | 0 | 24 | 0 | 6 |
| 14 | P | 75 | 8 | 0 | 7 |
| 15 | P | 90 | 8 | 0 | 5 |
| 16 | R | 0 | 8 | 0 | 9 |
| 17 | R | 45 | 8 | 0 | 8 |
| 18 | R | 75 | 8 | 0 | 7 |
| 19 | R | 90 | 8 | 0 | 7 |
| 20 | P | 45 | 8 | 2.5 | 1 |
| 21 | P | 75 | 8 | 2.5 | 5 |
| 22 | P | 90 | 8 | 2.5 | 5 |
| 23 | R | 0 | 8 | 2.5 | 5 |
| 24 | R | 45 | 8 | 2.5 | 7 |
| 25 | R | 75 | 8 | 2.5 | 1 |
| 26 | R | 90 | 8 | 2.5 | 1 |
| 30 | R | 0 | 8 | 5 | 2 |
| 41 | P | 45 | 16 | 0 | 3 |
| 42 | P | 75 | 16 | 0 | 8 |
| 43 | P | 90 | 16 | 0 | 8 |
| 44 | R | 0 | 16 | 0 | 9 |
| 45 | R | 45 | 16 | 0 | 7 |
| 46 | R | 75 | 16 | 0 | 7 |
| 47 | R | 90 | 16 | 0 | 7 |
| 51 | R | 0 | 16 | 2.5 | 1 |
| 69 | P | 45 | 24 | 0 | 5 |
| 70 | P | 75 | 24 | 0 | 5 |
| 71 | P | 90 | 24 | 0 | 6 |
| 72 | R | 0 | 24 | 0 | 4 |
| 73 | R | 45 | 24 | 0 | 4 |
| 74 | R | 75 | 24 | 0 | 4 |
| 75 | R | 90 | 24 | 0 | 4 |
| 97 | P | 33 | 8 | 0 | 1 |
| 98 | P | 33 | 8 | 2.5 | 2 |
| 99 | P | 33 | 8 | 5 | 2 |
| 100 | P | 33 | 8 | 15 | 3 |
| 101 | P | 33 | 16 | 0 | 2 |
| 102 | P | 33 | 16 | 2.5 | 2 |
| 103 | R | 33 | 8 | 0 | 3 |
| 104 | R | 33 | 8 | 2.5 | 1 |
| 105 | R | 33 | 8 | 5 | 3 |
| 106 | R | 33 | 8 | 15 | 3 |
| 109 | P | 60 | 8 | 0 | 1 |
| 110 | R | 60 | 8 | 0 | 1 |
| 111 | P | 60 | 8 | 2.5 | 1 |
| 113 | P | 60 | 8 | 5 | 1 |
| 115 | P | 60 | 8 | 15 | 1 |
| 117 | P | 60 | 16 | 0 | 1 |

**Notes.** *Col. 1*: orbit identifier in the `GalMer` database, unique for each set of initial orbital parameters, inclination and spin-orbit coupling of the experiments[a]. *Col. 2*: spin-orbit coupling. The orbit is prograde (P) if the *z*-component of the orbital spin is parallel to the *z* axis and retrograde (R) if anti-parallel. *Col. 3*: disc inclination of the second progenitor with respect to the orbital plane. The disc of the first progenitor is kept in the orbital plane when present. *Col. 4*: pericentre distance. *Col. 5*: initial orbital energy. *Col. 6*: total number of experiments with this set of orbital parameters in the sample of 202 dynamically relaxed remnants selected from the 1002 experiments available in the database (Sect. 3.3).

---

[a] Notice that the *orbit identifier* is different from the *orbit type* defined in the query sheet of the `GalMer` database.

Table A.2: Dynamical status of the 287 pre-selected models with apparently relaxed remnants hosting disc components

| | | Major mergers | | | | | | | Major mergers | | | | |
|---|---|---|---|---|---|---|---|---|---|---|---|---|---|
| No. | Model | $C_1$ | $C_2$ | $C_3$ | Merging | Dynam. | No. | Model | $C_1$ | $C_2$ | $C_3$ | Merging | Dynam. |
| (1) | (2) | (3) | (4) | (5) | (6) | (7) | (1) | (2) | (3) | (4) | (5) | (6) | (7) |
| 1 | gE0gSao1 | Y | Y | Y | SR | DR | 109 | gSagSdo24 | Y | Y | Y | SR | DR |
| 2 | gE0gSao5 | Y | Y | Y | SR | DR | 110 | gSagSdo30 | Y | Y | Y | SR | DR |
| 3 | gE0gSao13 | N | N | N | OM | – | 111 | gSagSdo41 | Y | Y | Y | SR | DR |
| 4 | gE0gSao16 | Y | Y | Y | SR | DR | 112 | gSagSdo42 | Y | Y | Y | SR | DR |
| 5 | gE0gSao17 | Y | Y | Y | SR | DR | 113 | gSagSdo43 | Y | Y | Y | SR | DR |
| ... | ... | ... | ... | ... | ... | ... | ... | ... | ... | ... | ... | ... | ... |

**Notes.** *Col. 1*: ordering number. *Col. 2*: model ID. *Cols. 3–5*: whether the remnant obeys criteria 1 to 3 imposed in Sect. 3.2 for major and minor mergers to be considered fully merged and structurally relaxed ("Y": yes; "N": no). *Col. 6*: merging status on the basis of the three criteria exposed in the previous three columns ("OM": on-going merger; "NR": non structurally relaxed; "SR": structurally relaxed). *Col. 7*: dynamical status of the structurally relaxed remnants ("DR": dynamically relaxed –virialised; "N-DR": non dynamically relaxed).

[A full version of this Table is available at: https://www.researchgate.net/publication/325905181_Formation_of_S0_galaxies_through_mergers_Morphological_properties_tidal_relics_lenses_ovals_and_other_inner_components_-_Version_of_the_corresponding_AA_paper_with_full_Appendices]

Table A.3: Orbital times and visual morphological type of the 204 fully merged and structurally relaxed remnants in our sample

| | | Major mergers | | | | | | | | | | |
|---|---|---|---|---|---|---|---|---|---|---|---|---|
| No. | Model | Morph. Type | $D_0$ [kpc] | $T_{per,1}$ [Gyr] | $D_{per,1}$ [kpc] | $T_{per,2}$ [Gyr] | $D_{per,2}$ [kpc] | $T_{fullmerger}$ [Gyr] | $D_{fullmerger}$ [kpc] | $T_{final}$ [Gyr] | $D_{final}$ [kpc] | $T_{relax}$ [Gyr] |
| (1) | (2) | (3) | (4) | (5) | (6) | (7) | (8) | (9) | (10) | (11) | (12) | (13) |
| 1 | gE0gSao1 | S0 | 99.98 | 0.35 | 11.41 | 0.75 | 0.77 | 0.75 | 0.768 | 3.00 | 0.272 | 2.25 |
| 2 | gE0gSao5 | S0 | 99.98 | 0.40 | 14.55 | 1.35 | 9.35 | 1.50 | 0.410 | 2.95 | 0.097 | 1.45 |
| 3 | gE0gSao16 | S0 | 99.98 | 0.35 | 11.89 | 0.85 | 8.94 | 1.00 | 0.755 | 3.00 | 0.067 | 2.00 |
| 4 | gE0gSao17 | E | 99.99 | 0.35 | 12.49 | 0.90 | 7.50 | 1.05 | 0.489 | 3.50 | 0.052 | 2.45 |
| 5 | gE0gSao18 | E | 99.99 | 0.35 | 12.47 | 0.95 | 8.22 | 1.05 | 0.945 | 3.50 | 0.066 | 2.45 |
| ... | ... | ... | ... | ... | ... | ... | ... | ... | ... | ... | ... | ... |

**Notes.** *Col. 1*: ordering number. *Col. 2*: model ID. *Col. 3-*: visual morphological type. *Col. 4*: initial distance between the mass centroids of the progenitors in kpc, $D_0$. *Col. 5*: time of the first pericentre passage in the orbit, $T_{per,1}$, in Gyr. *Col. 6*: distance between the mass centroids of the two progenitors in the first pericentre passage, $D_{per,1}$, in kpc. *Col. 7*: time of the second pericentre passage in the orbit, $T_{per,2}$, in Gyr. *Col. 8*: distance between centroids in the second pericentre passage, $D_{per,2}$, in kpc. *Col. 9*: time for the full merger of both galaxies, $T_{fullmerger}$, in Gyr. *Col. 10*: distance between mass centroids of the two progenitors in full merger, $D_{fullmerger}$, in kpc. *Col. 11*: total time computed in the model, $T_{final}$, in Gyr. *Col. 12*: distance between the centroids of the two progenitors in the final remnant, $D_{final}$, in kpc. *Col. 13*: total relaxation time experienced by the remnant after full merger, $T_{relax}$, in Gyr. The models marked with † are not dynamically relaxed (see Sect. 3.3).

[A full version of this Table is available at: https://www.researchgate.net/publication/325905181_Formation_of_S0_galaxies_through_mergers_Morphological_properties_tidal_relics_lenses_ovals_and_other_inner_components_-_Version_of_the_corresponding_AA_paper_with_full_Appendices]

Table A.4: Morphological features detected visually in the photometric images of the S0-like remnants

| No. | Model | Morph. Type | Merger-related features | Bar? | Global features | Tidal satellites? |
|---|---|---|---|---|---|---|
| (1) | (2) | (3) | (4) | (5) | (6) | (7) |
| 1 | gE0gSao1 | S0 | – | – | SL (oval), NB | Y |
| 2 | gE0gSao5 | S0 | Plume | – | SL, NSp | – |
| 3 | gE0gSao16 | S0 | – | – | SL | – |
| 4 | gE0gSao44 | S0 | – | – | LL | – |
| 5 | gE0gSbo5 | S0 | – | – | SL | – |
| ... | ... | ... | ... | ... | ... | ... |

**Notes.** *Col. 1*: ordering number. *Col. 2*: model ID. *Col. 3*: visual morphological type. *Col. 4*: morphological features detected in the photometric images of the remnant indicating past merger activity. Used acronyms: "DA": disc asymmetry; "TD": tidal debris; "TT": tidal tail. *Col. 5*: whether the final remnant is barred ("Y") or not ("–"). *Col. 6*: relevant morphological features detected in the photometric images of the remnant, not necessarily related to a merger origin. Used acronyms: "FL": flat lens; "ID": inner disc; " IID": inclined inner disc; "IPR": inner pseudo-ring; "IR": inner ring; "LL": lentil-shaped lens; "NB": nuclear bar; "ISp": inner spiral; "OR": outer ring; "OSR": outer spiral relics; "SL": spheroidal lens. *Col. 7*: whether there are tidal satellites orbiting the remnant within a FoV of 100×100 kpc² for the face-on and edge-on views ("Y": there is at least one; "–": none).

[A full version of this Table is available at: https://www.researchgate.net/publication/325905181_Formation_of_S0_galaxies_through_mergers_Morphological_properties_tidal_relics_lenses_ovals_and_other_inner_components_-_Version_of_the_corresponding_AA_paper_with_full_Appendices]

Table A.5: Statistics of morphological features in the major- and minor-merger S0-like samples

| Main disc morphological features[a] | Major mergers | | Minor mergers | |
|---|---|---|---|---|
| | No. | Percentage[b] | No. | Percentage[b] |
| Lenses (total) | 65 | 97% | 0 | ... |
| Spheroidal lenses | 14 | 21% | 0 | ... |
| Flat lenses | 43 | 64% | 0 | ... |
| Lentil-shaped lenses | 8 | 12% | 0 | ... |
| Ovals | 2 | 3% | 29 | 100% |
| Large bars[c] | 0 | ... | 29 | 100% |
| Warped main disc | 3 | 4% | 20 | 68% |
| Outer rings | 0 | ... | 1 | 3% |
| Spiral relics in outer disc | 5 | 7% | 23 | 79% |
| Inner/nuclear components[a] | Major mergers | | Minor mergers | |
| | No. | Percentage[b] | No. | Percentage[b] |
| Embedded co-planar inner discs | 16 | 24% | 0 | ... |
| Inclined inner discs | 14 | 21% | 0 | ... |
| Inner spirals | 12 | 18% | 0 | ... |
| Nuclear bars | 9 | 13% | 0 | ... |
| Inner rings | 2 | 3% | 0 | ... |
| Inner pseudo-rings | 5 | 7% | 1 | 3% |
| With any ICs | 39 | 58% | 1 | 3% |
| Merger-related features[a] | Major mergers | | Minor mergers | |
| | No. | Percentage[b] | No. | Percentage[b] |
| Tidal tails | 6 | 9% | 2 | 7% |
| Tidal debris | 10 | 15% | 10 | 34% |
| Plumes/ripples | 4 | 6% | 18 | 62% |
| Disc asymmetry | 1 | 1% | 6 | 21% |
| Shells | 0 | ... | 1 | 3% |
| Collisional rings | 1 | 1% | 0 | ... |
| With merger relics[d] | 16 | 24% | 19 | 66% |
| With no traces of past merger[d] | 51 | 76% | 10 | 34% |
| Miscellanea[a] | Major mergers | | Minor mergers | |
| | No. | Percentage[b] | No. | Percentage[b] |
| Tidal satellites | 32 | 48% | 0 | ... |

**Notes.** [a] We have distinguished between morphological features that can be directly attributed to past merger activity in the galaxy ("Merger-related") from those that are not strictly related to mergers or cannot be observationally attributed to them. Tidal satellites are provided in "Miscellanea" because, although they are produced by interactions and mergers by definition, it is difficult to establish the tidal nature of the satellites orbiting a galaxy just through visual inspection if the system is relaxed. We have preferred not considering them as merger relics (for more details, see Sect. 7.3).
[b] The percentages in the major- and minor-merger subsamples are computed independently for each one, considering a total of 67 S0-like remnants coming from major mergers and a total of 29 S0 ones from minor mergers.
[c] The gS0 progenitor of the minor merger experiments is strongly barred. There is only one S0 major-merger remnant with traces of a large bar (gSbgSbo72), but so diluted that we have not considered it as significant (see Sect. 7.1).
[d] The statistics for the S0-like remnants "with merger relics" include all remnants with one or more morphological features observable in the artificial photometric images pointing to their past merger origin. Those "with no traces of past merger" show none.

# Appendix B: Atlas of S0 and E/S0 remnants

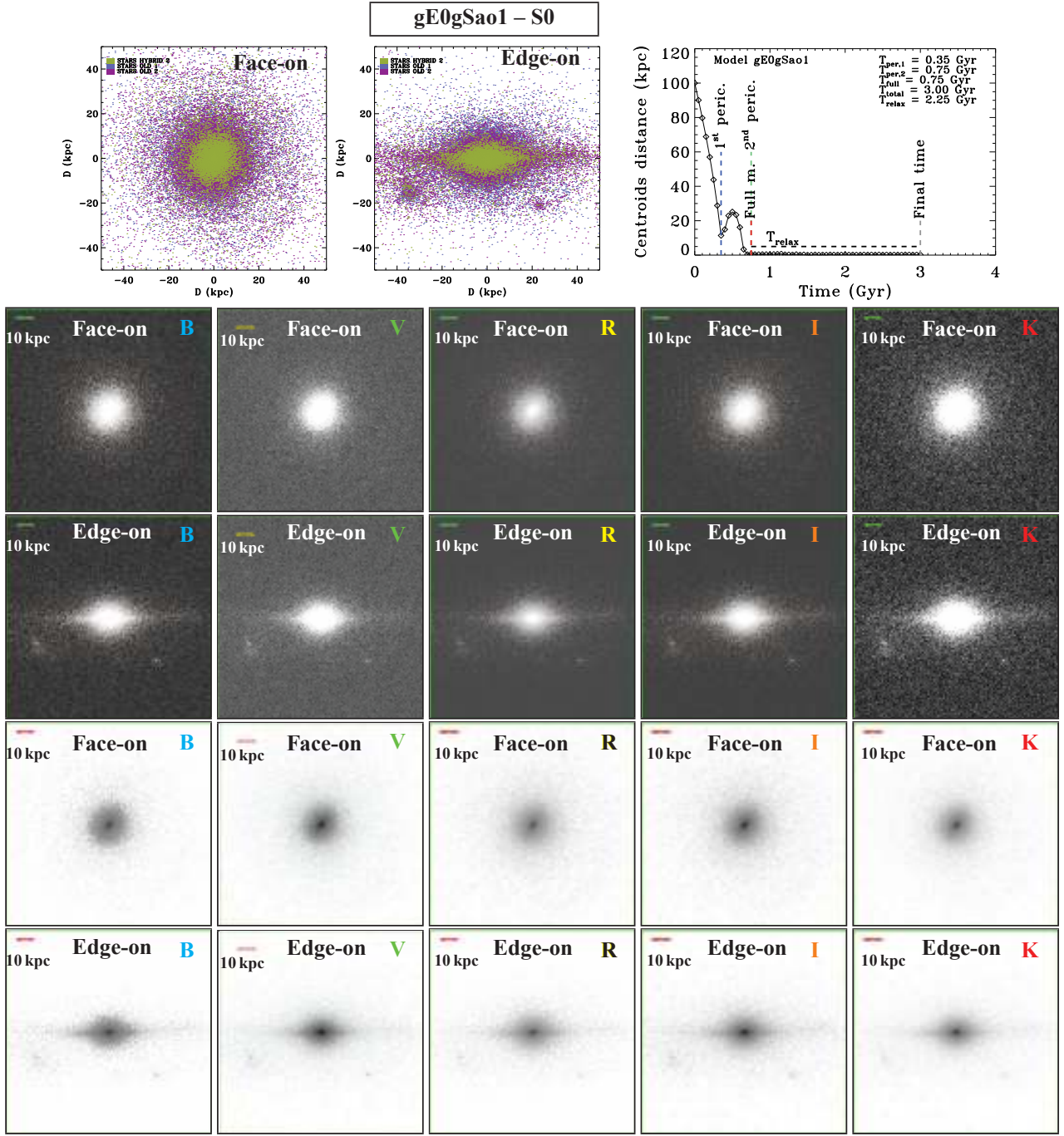

Fig. B.1: Morphology of the final remnant of model gE0gSao1 (S0). *Top left panels*: projected stellar particle-position maps for face-on and edge-on views of the remnant. The particles have been coloured according to their original progenitor (*dark green*: hybrid particles, progenitor 1; *blue*: old stars, progenitor 1; *light green*: hybrid particles, progenitor 2; *purple*: old stars, progenitor 2). *Top right panel*: time evolution of the distance between the centroids of the two merging galaxies during the simulation, with the relevant orbital times indicated. *Main set of panels*: mock photometric images in the $B$, $V$, $R$, $I$, and $K$ bands of the final remnant for face-on and edge-on views, assuming a distance of 30 Mpc. In the first two rows, we use a logarithmic grayscale that highlights the structure in the outskirts of the galaxy. In the two bottom rows, we use a reversed scale to enhance the sub-structures in the core instead. The FoV in all frames is $100 \times 100$ kpc$^2$. The total extension of the outer disc is difficult to appreciate in the face-on views due to its grainy appearance, but it typically extends beyond approximately two to four times the outer radius of the central lens or oval, as seen in the corresponding edge-on views.

[**Note.**– A complete version of the atlas with the 96 S0 and E/S0 remnants is available at: `https://www.researchgate.net/publication/325905181_Formation_of_S0_galaxies_through_mergers_Morphological_properties_tidal_relics_lenses_ovals_and_other_inner_components_-_Version_of_the_corresponding_AA_paper_with_full_Appendices`]